\documentclass{article}

\usepackage{microtype}
\usepackage{graphicx}
\usepackage{booktabs}
\usepackage{hyperref}

\usepackage[preprint]{icml2026}
\hypersetup{pdfsubject={Research preprint (provisional ICML formatting)}}

\usepackage[utf8]{inputenc}
\usepackage[T1]{fontenc}
\usepackage{url}
\usepackage{xurl} 
\usepackage{amsfonts}
\usepackage{amsmath}
\usepackage{amssymb}
\usepackage{nicefrac}
\usepackage{xcolor}
\usepackage{multirow}
\usepackage{longtable}
\usepackage{float}
\usepackage{placeins}

\newcommand{\R}{\ensuremath{R}}
\newcommand{\dww}{\ensuremath{\Delta_{\mathrm{WW}}}}
\newcommand{\dse}{\ensuremath{\Delta_{\mathrm{SE}}}}
\newcommand{\ci}[2]{\,(#1,\,#2)}
\newcommand{\TableCohort}{%
\begin{table}[H]
\centering
\caption{Cohort flow (recordings; persons in parentheses). Reconstructed from the original eligibility inventories (metadata only). Primary night: lowest-numbered eligible night per person (VAL, TEST); EXT used all eligible nights with the person as the bootstrap cluster; TRAIN used all eligible nights.}
\label{tab:cohort}
\footnotesize
\setlength{\tabcolsep}{4pt}
\begin{tabular}{@{}lrr@{}}
\toprule
 & Sleep Cassette (SC) & Sleep Telemetry (ST) \\
\midrule
Source & 153 (78) & 44 (22) \\
Lights-off outside signal & $-$3 & $-$3 \\
Scoring gap/overlap before first N2 & $-$1 & $-$13 \\
N2 active at lights-off & $-$5 & 0 \\
Not awake (W) at lights-off & $-$3 & 0 \\
Analysable & 141 (76) & 28 (17) \\
\midrule
TRAIN (development) & 84 (46) & --- \\
VAL (primary and sensitivity) & 15 + 12 (15) & --- \\
TEST (primary and sensitivity) & 15 + 15 (15) & --- \\
EXT (all eligible nights) & --- & 28 (17) \\
\bottomrule
\end{tabular}
\end{table}}

\newcommand{\TableFrozen}{%
\begin{table}[t]
\centering
\caption{Prespecified held-out comparisons for prediction of the first Stage~2 epoch. Window-weighted Brier; $R$ in \%, with the original 95\% participant-cluster intervals from each evaluation. $\Delta_{\mathrm{WW}}$: absolute Brier difference. Sensitivity rows pool all eligible nights (VAL 27, TEST 30) and are descriptive. The external CNN comparison was not formally tested because the primary comparison did not meet its prespecified criteria.}
\label{tab:frozen}
\small
\setlength{\tabcolsep}{4pt}
\resizebox{\linewidth}{!}{%
\begin{tabular}{@{}lccc@{}}
\toprule
 & VAL & TEST & EXT \\
\midrule
Persons / nights; windows (positive) & 15 / 15; 905 (147) & 15 / 15; 809 (150) & 17 / 28; 1438 (280) \\
Brier: Constant / SpecLR-18 & 0.137 / 0.106 & 0.153 / 0.125 & 0.160 / 0.154 \\
Brier: SpecLR-46 / RawCNN-46 & --- & 0.118 / 0.095 & 0.149 / 0.121 \\
\midrule
Spectral model vs constant, $R$ & 22.3 (14.6, 30.0) & 18.3 (8.8, 28.8) & 3.4 ($-$0.9, 7.6) \\
\quad absolute $\Delta_{\mathrm{WW}}$ & 0.0305 & 0.0280 & 0.0055 \\

CNN vs SpecLR-18, $R$ & --- & 23.8 (4.7, 41.2), criteria met & 21.3 (10.5, 30.3), not tested \\
CNN vs SpecLR-46 (descriptive) & --- & 19.1 ($-$0.4, 36.4) & 18.5 (8.2, 26.7) \\
\midrule
All nights: spectral vs constant & 19.3 (12.9, 25.5) & 12.0 (0.4, 21.8) & (all nights primary) \\
All nights: CNN vs SpecLR-18 / SpecLR-46 & --- & 24.9 (9.0, 36.8) / 18.5 (1.5, 30.9) & --- \\
\bottomrule
\end{tabular}}
\end{table}}

\newcommand{\TableModels}{%
\begin{table}[H]
\centering
\caption{Metrics of EEG forecasters in the internal and external test cohorts , generated from the original evaluation records. WW: window-weighted; SE: subject-equal; CIL: calibration-in-the-large. Calibration slope is undefined for a constant predictor (n/a; the TEST result file records a numerical value of 0.626 for the constant, which is not meaningful). Time-matched AUROC as defined in the original evaluation protocol: cross-person pairs within elapsed-time strata.}
\label{tab:models}
\footnotesize
\setlength{\tabcolsep}{3.5pt}
\resizebox{\linewidth}{!}{%
\begin{tabular}{@{}llccccccc@{}}
\toprule
Set & Model & Brier WW & Brier SE & Log loss WW & CIL & Slope & Within-rec.\ AUROC & Time-matched AUROC \\
\midrule
\multirow{4}{*}{TEST} & Constant & 0.153 & 0.263 & 0.487 & $-$0.045 & n/a & 0.50 & 0.50 \\
 & SpecLR-18 & 0.125 & 0.214 & 0.403 & $-$0.033 & 1.39 & 0.80 & 0.78 \\
 & SpecLR-46 & 0.118 & 0.196 & 0.387 & $-$0.025 & 1.81 & 0.84 & 0.81 \\
 & RawCNN-46 & 0.095 & 0.136 & 0.324 & $+$0.020 & 1.45 & 0.88 & 0.90 \\
\midrule
\multirow{4}{*}{EXT} & Constant & 0.160 & 0.207 & 0.504 & $-$0.054 & n/a & 0.50 & 0.50 \\
 & SpecLR-18 & 0.154 & 0.208 & 0.499 & $-$0.130 & 1.33 & 0.82 & 0.80 \\
 & SpecLR-46 & 0.149 & 0.200 & 0.473 & $-$0.124 & 1.42 & 0.83 & 0.79 \\
 & RawCNN-46 & 0.121 & 0.153 & 0.394 & $+$0.006 & 1.04 & 0.79 & 0.78 \\
\bottomrule
\end{tabular}}
\end{table}}

\newcommand{\TableCentral}{%
\begin{table}[t]
\centering
\caption{Held-out comparisons against development-fitted scored-stage references (retrospective analysis). $R$ in \% (95\% person-cluster re-bootstrap interval) for the second model relative to the first. Stage-only and Stage+time use the retrospectively scored stage of the input epoch and are diagnostic references, not deployable models. Stage-reference improvements do not depend on the EEG model. EEG-versus-constant intervals here are recomputed from the original predictions (the original intervals are in Table~\ref{tab:frozen}). Bold: interval excludes zero.}
\label{tab:central}
\footnotesize
\setlength{\tabcolsep}{2.6pt}
\resizebox{\linewidth}{!}{%
\begin{tabular}{@{}lccccc@{}}
\toprule
 & VAL & \multicolumn{2}{c}{TEST} & \multicolumn{2}{c}{EXT} \\
\cmidrule(lr){3-4}\cmidrule(lr){5-6}
 & SpecLR-18 & SpecLR-18 & RawCNN-46 & SpecLR-18 & RawCNN-46 \\
\midrule
EEG vs constant & \textbf{22.3 (14.7, 29.8)} & \textbf{18.3 (8.8, 28.8)} & \textbf{37.7 (16.1, 55.3)} & 3.4 ($-$1.0, 7.6) & \textbf{24.0 (15.8, 31.8)} \\
Stage-only vs constant & \textbf{15.5 (4.5, 26.3)} & \textbf{26.3 (15.6, 37.4)} & $\leftarrow$ & \textbf{29.5 (22.6, 36.3)} & $\leftarrow$ \\
Stage+time vs constant & 12.8 ($-$1.8, 26.4) & \textbf{30.3 (18.4, 42.3)} & $\leftarrow$ & \textbf{32.2 (24.9, 39.4)} & $\leftarrow$ \\
EEG vs Stage-only & 8.1 ($-$9.1, 22.1) & $-$10.9 ($-$26.9, 1.4) & 15.5 ($-$5.2, 32.9) & \textbf{$-$37.0 ($-$51.5, $-$24.4)} & $-$7.8 ($-$18.8, 2.6) \\
EEG vs Stage+time & 10.9 ($-$8.9, 26.8) & \textbf{$-$17.1 ($-$39.3, $-$0.6)} & 10.7 ($-$16.0, 30.6) & \textbf{$-$42.6 ($-$58.6, $-$28.8)} & \textbf{$-$12.2 ($-$25.3, $-$0.4)} \\
EEG vs Stage+time, N1 & 7.6 ($-$40.8, 33.3) & $-$26.8 ($-$77.5, 3.5) & \textbf{28.5 (7.0, 43.3)} & \textbf{$-$85.5 ($-$132.9, $-$52.9)} & $-$15.5 ($-$36.9, 0.9) \\
EEG vs Stage+time, W & \textbf{14.6 (5.2, 25.8)} & $-$6.3 ($-$32.1, 5.8) & $-$9.2 ($-$72.8, 21.6) & 0.6 ($-$4.2, 5.0) & $-$8.9 ($-$28.6, 1.8) \\
\bottomrule
\end{tabular}}
\end{table}}

\newcommand{\TableRtwoFull}{%
{\footnotesize\setlength{\tabcolsep}{2.5pt}
\begin{longtable}{@{}lllrrrrl@{}}
\caption{Stage-stratified evaluation of held-out forecasts (retrospective, descriptive). Brier reduction of each previously evaluated model relative to the constant reference on the same windows. $\Delta_{\mathrm{WW}}$: absolute window-weighted Brier difference; $R$: relative reduction (\%) with 95\% person-cluster CI (n.r.: fewer than 8 persons, CI not reported). Subsets are scorer-defined and were not available to the models.}\label{tab:r2full}\\
\toprule
Set & Model & Subset & Windows (pos.) & Persons & Recs & $\Delta_{\mathrm{WW}}$ & $R$ (\%) \\
\midrule
\endfirsthead
\multicolumn{8}{@{}l}{\emph{(continued)}}\\
\toprule
Set & Model & Subset & Windows (pos.) & Persons & Recs & $\Delta_{\mathrm{WW}}$ & $R$ (\%) \\
\midrule
\endhead
VAL & SpecLR-18 & All & 905 (147) & 15 & 15 & 0.0305 & 22.3 (14.4, 29.9) \\
VAL & SpecLR-18 & Stage 1 & 214 (91) & 13 & 13 & 0.0809 & 24.9 (16.5, 33.9) \\
VAL & SpecLR-18 & W & 691 (56) & 15 & 15 & 0.0149 & 19.1 (5.4, 29.4) \\
VAL & SpecLR-18 & Pre-sleep & 615 (40) & 15 & 15 & 0.0155 & 23.3 (7.1, 35.1) \\
VAL & SpecLR-18 & Returning W & 76 (16) & 9 & 9 & 0.0098 & 5.7 ($-$6.5, 18.8) \\
\addlinespace[1pt]
TEST & SpecLR-18 & All & 809 (150) & 15 & 15 & 0.0280 & 18.3 (8.8, 29.0) \\
TEST & SpecLR-18 & Stage 1 & 200 (106) & 14 & 14 & 0.1119 & 27.9 (16.2, 39.7) \\
TEST & SpecLR-18 & W & 609 (44) & 15 & 15 & 0.0005 & 0.7 ($-$8.5, 13.1) \\
TEST & SpecLR-18 & Pre-sleep & 518 (33) & 15 & 15 & 0.0022 & 3.4 ($-$7.5, 21.0) \\
TEST & SpecLR-18 & Returning W & 91 (11) & 6 & 6 & $-$0.0092 & $-$8.7 (n.r.) \\
\addlinespace[1pt]
TEST & SpecLR-46 & All & 809 (150) & 15 & 15 & 0.0353 & 23.1 (13.8, 33.2) \\
TEST & SpecLR-46 & Stage 1 & 200 (106) & 14 & 14 & 0.1407 & 35.1 (26.1, 44.4) \\
TEST & SpecLR-46 & W & 609 (44) & 15 & 15 & 0.0007 & 1.0 ($-$7.1, 13.0) \\
TEST & SpecLR-46 & Pre-sleep & 518 (33) & 15 & 15 & 0.0014 & 2.2 ($-$7.5, 18.3) \\
TEST & SpecLR-46 & Returning W & 91 (11) & 6 & 6 & $-$0.0034 & $-$3.2 (n.r.) \\
\addlinespace[1pt]
TEST & RawCNN-46 & All & 809 (150) & 15 & 15 & 0.0578 & 37.7 (16.7, 55.6) \\
TEST & RawCNN-46 & Stage 1 & 200 (106) & 14 & 14 & 0.2378 & 59.3 (46.1, 72.0) \\
TEST & RawCNN-46 & W & 609 (44) & 15 & 15 & $-$0.0014 & $-$1.9 ($-$38.1, 25.2) \\
TEST & RawCNN-46 & Pre-sleep & 518 (33) & 15 & 15 & $-$0.0012 & $-$1.9 ($-$45.8, 26.1) \\
TEST & RawCNN-46 & Returning W & 91 (11) & 6 & 6 & $-$0.0021 & $-$2.0 (n.r.) \\
\addlinespace[1pt]
EXT & SpecLR-18 & All & 1438 (280) & 17 & 28 & 0.0055 & 3.4 ($-$0.7, 7.6) \\
EXT & SpecLR-18 & Stage 1 & 331 (196) & 17 & 28 & 0.0083 & 1.9 ($-$2.3, 5.7) \\
EXT & SpecLR-18 & W & 1107 (84) & 17 & 28 & 0.0046 & 6.2 ($-$2.7, 16.8) \\
EXT & SpecLR-18 & Pre-sleep & 946 (48) & 17 & 28 & 0.0094 & 16.7 (2.1, 39.5) \\
EXT & SpecLR-18 & Returning W & 161 (36) & 13 & 17 & $-$0.0237 & $-$13.1 ($-$18.1, $-$5.3) \\
\addlinespace[1pt]
EXT & SpecLR-46 & All & 1438 (280) & 17 & 28 & 0.0107 & 6.7 (2.5, 10.5) \\
EXT & SpecLR-46 & Stage 1 & 331 (196) & 17 & 28 & 0.0295 & 6.6 (3.2, 9.6) \\
EXT & SpecLR-46 & W & 1107 (84) & 17 & 28 & 0.0051 & 6.8 ($-$1.9, 17.4) \\
EXT & SpecLR-46 & Pre-sleep & 946 (48) & 17 & 28 & 0.0095 & 16.9 (2.4, 40.5) \\
EXT & SpecLR-46 & Returning W & 161 (36) & 13 & 17 & $-$0.0210 & $-$11.6 ($-$17.7, $-$3.1) \\
\addlinespace[1pt]
EXT & RawCNN-46 & All & 1438 (280) & 17 & 28 & 0.0383 & 24.0 (15.7, 31.8) \\
EXT & RawCNN-46 & Stage 1 & 331 (196) & 17 & 28 & 0.1732 & 38.9 (27.8, 49.8) \\
EXT & RawCNN-46 & W & 1107 (84) & 17 & 28 & $-$0.0021 & $-$2.8 ($-$11.7, 4.8) \\
EXT & RawCNN-46 & Pre-sleep & 946 (48) & 17 & 28 & $-$0.0034 & $-$6.1 ($-$21.1, 6.5) \\
EXT & RawCNN-46 & Returning W & 161 (36) & 13 & 17 & 0.0061 & 3.4 ($-$19.4, 16.8) \\
\bottomrule
\end{longtable}}}

\newcommand{\TableRthreeB}{%
{\footnotesize\setlength{\tabcolsep}{2.5pt}
\begin{longtable}{@{}lllrrl@{}}
\caption{Stage-conditioned comparisons on previously held-out forecasts (retrospective analysis; participant-cluster bootstrap intervals, 10{,}000 person-cluster replicates, seed 20261026; n.r.: fewer than 8 persons). Subscript 18: reference refitted on the 18-person SpecLR-18 training roster.}\label{tab:r3bfull}\\
\toprule
Set & Stratum & Contrast & Persons & $\Delta_{\mathrm{WW}}$ & $R$ (\%) \\
\midrule
\endfirsthead
\multicolumn{6}{@{}l}{\emph{(continued)}}\\
\toprule
Set & Stratum & Contrast & Persons & $\Delta_{\mathrm{WW}}$ & $R$ (\%) \\
\midrule
\endhead
VAL & All & Constant $\to$ SpecLR-18 & 15 & 0.0305 & 22.3 (14.7, 29.8) \\
VAL & All & Constant $\to$ Stage-only & 15 & 0.0211 & 15.5 (4.5, 26.3) \\
VAL & All & Constant $\to$ Stage+time & 15 & 0.0175 & 12.8 ($-$1.8, 26.4) \\
VAL & All & Stage-only $\to$ SpecLR-18 & 15 & 0.0094 & 8.1 ($-$9.1, 22.1) \\
VAL & All & Stage+time $\to$ SpecLR-18 & 15 & 0.0130 & 10.9 ($-$8.9, 26.8) \\
VAL & All & Stage-only$_{18}$ $\to$ SpecLR-18 & 15 & 0.0104 & 8.9 ($-$9.0, 23.4) \\
VAL & All & Stage+time$_{18}$ $\to$ SpecLR-18 & 15 & 0.0101 & 8.7 ($-$8.4, 22.7) \\
\addlinespace[1pt]
VAL & Stage 1 & Constant $\to$ SpecLR-18 & 13 & 0.0809 & 24.9 (16.5, 33.9) \\
VAL & Stage 1 & Constant $\to$ Stage-only & 13 & 0.0783 & 24.1 (6.0, 42.9) \\
VAL & Stage 1 & Constant $\to$ Stage+time & 13 & 0.0607 & 18.7 ($-$5.8, 44.2) \\
VAL & Stage 1 & Stage-only $\to$ SpecLR-18 & 13 & 0.0026 & 1.1 ($-$38.1, 25.7) \\
VAL & Stage 1 & Stage+time $\to$ SpecLR-18 & 13 & 0.0202 & 7.6 ($-$40.8, 33.3) \\
\addlinespace[1pt]
VAL & W & Constant $\to$ SpecLR-18 & 15 & 0.0149 & 19.1 (5.7, 29.5) \\
VAL & W & Constant $\to$ Stage-only & 15 & 0.0034 & 4.4 ($-$8.5, 17.8) \\
VAL & W & Constant $\to$ Stage+time & 15 & 0.0041 & 5.2 ($-$6.9, 17.8) \\
VAL & W & Stage-only $\to$ SpecLR-18 & 15 & 0.0115 & 15.4 (5.8, 26.3) \\
VAL & W & Stage+time $\to$ SpecLR-18 & 15 & 0.0108 & 14.6 (5.2, 25.8) \\
\addlinespace[1pt]
VAL & Pre-sleep & Constant $\to$ SpecLR-18 & 15 & 0.0155 & 23.3 (7.7, 35.2) \\
VAL & Pre-sleep & Constant $\to$ Stage-only & 15 & 0.0057 & 8.5 ($-$8.0, 26.1) \\
VAL & Pre-sleep & Constant $\to$ Stage+time & 15 & 0.0063 & 9.4 ($-$6.1, 25.3) \\
VAL & Pre-sleep & Stage-only $\to$ SpecLR-18 & 15 & 0.0098 & 16.2 (3.2, 29.4) \\
VAL & Pre-sleep & Stage+time $\to$ SpecLR-18 & 15 & 0.0092 & 15.3 (3.1, 29.2) \\
\addlinespace[1pt]
VAL & Returning W & Constant $\to$ SpecLR-18 & 9 & 0.0098 & 5.7 ($-$6.7, 18.2) \\
VAL & Returning W & Constant $\to$ Stage-only & 9 & $-$0.0149 & $-$8.7 ($-$12.7, $-$2.0) \\
VAL & Returning W & Constant $\to$ Stage+time & 9 & $-$0.0139 & $-$8.2 ($-$10.7, $-$2.1) \\
VAL & Returning W & Stage-only $\to$ SpecLR-18 & 9 & 0.0247 & 13.3 (1.1, 24.4) \\
VAL & Returning W & Stage+time $\to$ SpecLR-18 & 9 & 0.0237 & 12.8 ($-$0.4, 24.3) \\
\addlinespace[1pt]
TEST & All & Constant $\to$ SpecLR-18 & 15 & 0.0280 & 18.3 (8.8, 28.8) \\
TEST & All & Constant $\to$ SpecLR-46 & 15 & 0.0353 & 23.1 (13.8, 32.9) \\
TEST & All & Constant $\to$ RawCNN-46 & 15 & 0.0578 & 37.7 (16.1, 55.3) \\
TEST & All & Constant $\to$ Stage-only & 15 & 0.0403 & 26.3 (15.6, 37.4) \\
TEST & All & Constant $\to$ Stage+time & 15 & 0.0463 & 30.3 (18.4, 42.3) \\
TEST & All & Stage-only $\to$ SpecLR-18 & 15 & $-$0.0123 & $-$10.9 ($-$26.9, 1.4) \\
TEST & All & Stage+time $\to$ SpecLR-18 & 15 & $-$0.0183 & $-$17.1 ($-$39.3, $-$0.6) \\
TEST & All & Stage-only$_{18}$ $\to$ SpecLR-18 & 15 & $-$0.0125 & $-$11.1 ($-$28.3, 1.8) \\
TEST & All & Stage+time$_{18}$ $\to$ SpecLR-18 & 15 & $-$0.0128 & $-$11.4 ($-$27.7, 1.1) \\
TEST & All & Stage-only $\to$ SpecLR-46 & 15 & $-$0.0050 & $-$4.4 ($-$16.2, 5.3) \\
TEST & All & Stage+time $\to$ SpecLR-46 & 15 & $-$0.0110 & $-$10.3 ($-$27.5, 3.2) \\
TEST & All & Stage-only $\to$ RawCNN-46 & 15 & 0.0174 & 15.5 ($-$5.2, 32.9) \\
TEST & All & Stage+time $\to$ RawCNN-46 & 15 & 0.0114 & 10.7 ($-$16.0, 30.6) \\
TEST & All & SpecLR-46 $\to$ RawCNN-46 & 15 & 0.0224 & 19.1 ($-$1.0, 36.4) \\
\addlinespace[1pt]
TEST & Stage 1 & Constant $\to$ SpecLR-18 & 14 & 0.1119 & 27.9 (16.1, 39.7) \\
TEST & Stage 1 & Constant $\to$ SpecLR-46 & 14 & 0.1407 & 35.1 (26.0, 44.5) \\
TEST & Stage 1 & Constant $\to$ RawCNN-46 & 14 & 0.2378 & 59.3 (46.0, 71.8) \\
TEST & Stage 1 & Constant $\to$ Stage-only & 14 & 0.1489 & 37.1 (20.4, 53.0) \\
TEST & Stage 1 & Constant $\to$ Stage+time & 14 & 0.1730 & 43.2 (25.4, 60.1) \\
TEST & Stage 1 & Stage-only $\to$ SpecLR-18 & 14 & $-$0.0370 & $-$14.7 ($-$51.5, 9.3) \\
TEST & Stage 1 & Stage+time $\to$ SpecLR-18 & 14 & $-$0.0611 & $-$26.8 ($-$77.5, 3.5) \\
TEST & Stage 1 & Stage-only $\to$ SpecLR-46 & 14 & $-$0.0082 & $-$3.2 ($-$30.1, 14.3) \\
TEST & Stage 1 & Stage+time $\to$ SpecLR-46 & 14 & $-$0.0323 & $-$14.2 ($-$53.0, 8.7) \\
TEST & Stage 1 & Stage-only $\to$ RawCNN-46 & 14 & 0.0890 & 35.3 (21.5, 47.0) \\
TEST & Stage 1 & Stage+time $\to$ RawCNN-46 & 14 & 0.0648 & 28.5 (7.0, 43.3) \\
TEST & Stage 1 & SpecLR-46 $\to$ RawCNN-46 & 14 & 0.0971 & 37.4 (22.3, 52.5) \\
\addlinespace[1pt]
TEST & W & Constant $\to$ SpecLR-18 & 15 & 0.0005 & 0.7 ($-$8.6, 12.9) \\
TEST & W & Constant $\to$ SpecLR-46 & 15 & 0.0007 & 1.0 ($-$7.1, 12.7) \\
TEST & W & Constant $\to$ RawCNN-46 & 15 & $-$0.0014 & $-$1.9 ($-$39.3, 25.2) \\
TEST & W & Constant $\to$ Stage-only & 15 & 0.0047 & 6.5 ($-$4.5, 22.9) \\
TEST & W & Constant $\to$ Stage+time & 15 & 0.0048 & 6.6 ($-$3.5, 22.8) \\
TEST & W & Stage-only $\to$ SpecLR-18 & 15 & $-$0.0042 & $-$6.2 ($-$30.7, 5.3) \\
TEST & W & Stage+time $\to$ SpecLR-18 & 15 & $-$0.0042 & $-$6.3 ($-$32.1, 5.8) \\
TEST & W & Stage-only $\to$ SpecLR-46 & 15 & $-$0.0040 & $-$5.9 ($-$28.5, 5.9) \\
TEST & W & Stage+time $\to$ SpecLR-46 & 15 & $-$0.0041 & $-$6.1 ($-$29.6, 6.1) \\
TEST & W & Stage-only $\to$ RawCNN-46 & 15 & $-$0.0060 & $-$9.0 ($-$71.6, 21.5) \\
TEST & W & Stage+time $\to$ RawCNN-46 & 15 & $-$0.0061 & $-$9.2 ($-$72.8, 21.6) \\
TEST & W & SpecLR-46 $\to$ RawCNN-46 & 15 & $-$0.0021 & $-$2.9 ($-$35.2, 18.9) \\
\addlinespace[1pt]
TEST & Pre-sleep & Constant $\to$ SpecLR-18 & 15 & 0.0022 & 3.4 ($-$7.7, 21.0) \\
TEST & Pre-sleep & Constant $\to$ SpecLR-46 & 15 & 0.0014 & 2.2 ($-$7.6, 18.3) \\
TEST & Pre-sleep & Constant $\to$ RawCNN-46 & 15 & $-$0.0012 & $-$1.9 ($-$46.0, 26.1) \\
TEST & Pre-sleep & Constant $\to$ Stage-only & 15 & 0.0059 & 9.0 ($-$4.1, 36.0) \\
TEST & Pre-sleep & Constant $\to$ Stage+time & 15 & 0.0062 & 9.4 ($-$2.6, 35.0) \\
TEST & Pre-sleep & Stage-only $\to$ SpecLR-18 & 15 & $-$0.0037 & $-$6.1 ($-$41.5, 4.5) \\
TEST & Pre-sleep & Stage+time $\to$ SpecLR-18 & 15 & $-$0.0039 & $-$6.6 ($-$41.0, 4.3) \\
TEST & Pre-sleep & Stage-only $\to$ SpecLR-46 & 15 & $-$0.0045 & $-$7.5 ($-$43.0, 3.7) \\
TEST & Pre-sleep & Stage+time $\to$ SpecLR-46 & 15 & $-$0.0047 & $-$8.0 ($-$42.4, 3.3) \\
TEST & Pre-sleep & Stage-only $\to$ RawCNN-46 & 15 & $-$0.0071 & $-$11.9 ($-$108.2, 19.6) \\
TEST & Pre-sleep & Stage+time $\to$ RawCNN-46 & 15 & $-$0.0074 & $-$12.4 ($-$107.4, 19.3) \\
TEST & Pre-sleep & SpecLR-46 $\to$ RawCNN-46 & 15 & $-$0.0027 & $-$4.1 ($-$47.8, 18.8) \\
\addlinespace[1pt]
TEST & Returning W & Constant $\to$ SpecLR-18 & 6 & $-$0.0092 & $-$8.7 (n.r.) \\
TEST & Returning W & Constant $\to$ SpecLR-46 & 6 & $-$0.0034 & $-$3.2 (n.r.) \\
TEST & Returning W & Constant $\to$ RawCNN-46 & 6 & $-$0.0021 & $-$2.0 (n.r.) \\
TEST & Returning W & Constant $\to$ Stage-only & 6 & $-$0.0022 & $-$2.1 (n.r.) \\
TEST & Returning W & Constant $\to$ Stage+time & 6 & $-$0.0032 & $-$3.0 (n.r.) \\
TEST & Returning W & Stage-only $\to$ SpecLR-18 & 6 & $-$0.0070 & $-$6.5 (n.r.) \\
TEST & Returning W & Stage+time $\to$ SpecLR-18 & 6 & $-$0.0060 & $-$5.5 (n.r.) \\
TEST & Returning W & Stage-only $\to$ SpecLR-46 & 6 & $-$0.0012 & $-$1.1 (n.r.) \\
TEST & Returning W & Stage+time $\to$ SpecLR-46 & 6 & $-$0.0002 & $-$0.2 (n.r.) \\
TEST & Returning W & Stage-only $\to$ RawCNN-46 & 6 & 0.0001 & 0.1 (n.r.) \\
TEST & Returning W & Stage+time $\to$ RawCNN-46 & 6 & 0.0011 & 1.0 (n.r.) \\
TEST & Returning W & SpecLR-46 $\to$ RawCNN-46 & 6 & 0.0013 & 1.1 (n.r.) \\
\addlinespace[1pt]
EXT & All & Constant $\to$ SpecLR-18 & 17 & 0.0055 & 3.4 ($-$1.0, 7.6) \\
EXT & All & Constant $\to$ SpecLR-46 & 17 & 0.0107 & 6.7 (2.4, 10.6) \\
EXT & All & Constant $\to$ RawCNN-46 & 17 & 0.0383 & 24.0 (15.8, 31.8) \\
EXT & All & Constant $\to$ Stage-only & 17 & 0.0471 & 29.5 (22.6, 36.3) \\
EXT & All & Constant $\to$ Stage+time & 17 & 0.0515 & 32.2 (24.9, 39.4) \\
EXT & All & Stage-only $\to$ SpecLR-18 & 17 & $-$0.0416 & $-$37.0 ($-$51.5, $-$24.4) \\
EXT & All & Stage+time $\to$ SpecLR-18 & 17 & $-$0.0461 & $-$42.6 ($-$58.6, $-$28.8) \\
EXT & All & Stage-only$_{18}$ $\to$ SpecLR-18 & 17 & $-$0.0423 & $-$37.8 ($-$53.6, $-$24.3) \\
EXT & All & Stage+time$_{18}$ $\to$ SpecLR-18 & 17 & $-$0.0406 & $-$35.7 ($-$49.8, $-$23.6) \\
EXT & All & Stage-only $\to$ SpecLR-46 & 17 & $-$0.0364 & $-$32.3 ($-$46.8, $-$19.9) \\
EXT & All & Stage+time $\to$ SpecLR-46 & 17 & $-$0.0408 & $-$37.7 ($-$54.1, $-$23.8) \\
EXT & All & Stage-only $\to$ RawCNN-46 & 17 & $-$0.0088 & $-$7.8 ($-$18.8, 2.6) \\
EXT & All & Stage+time $\to$ RawCNN-46 & 17 & $-$0.0132 & $-$12.2 ($-$25.3, $-$0.4) \\
EXT & All & SpecLR-46 $\to$ RawCNN-46 & 17 & 0.0276 & 18.5 (8.4, 26.8) \\
\addlinespace[1pt]
EXT & Stage 1 & Constant $\to$ SpecLR-18 & 17 & 0.0083 & 1.9 ($-$2.3, 5.7) \\
EXT & Stage 1 & Constant $\to$ SpecLR-46 & 17 & 0.0295 & 6.6 (3.2, 9.8) \\
EXT & Stage 1 & Constant $\to$ RawCNN-46 & 17 & 0.1732 & 38.9 (27.8, 49.7) \\
EXT & Stage 1 & Constant $\to$ Stage-only & 17 & 0.1907 & 42.8 (31.1, 53.4) \\
EXT & Stage 1 & Constant $\to$ Stage+time & 17 & 0.2097 & 47.1 (35.4, 58.0) \\
EXT & Stage 1 & Stage-only $\to$ SpecLR-18 & 17 & $-$0.1824 & $-$71.6 ($-$110.3, $-$42.8) \\
EXT & Stage 1 & Stage+time $\to$ SpecLR-18 & 17 & $-$0.2014 & $-$85.5 ($-$132.9, $-$52.9) \\
EXT & Stage 1 & Stage-only $\to$ SpecLR-46 & 17 & $-$0.1613 & $-$63.3 ($-$100.7, $-$35.7) \\
EXT & Stage 1 & Stage+time $\to$ SpecLR-46 & 17 & $-$0.1802 & $-$76.5 ($-$122.2, $-$44.8) \\
EXT & Stage 1 & Stage-only $\to$ RawCNN-46 & 17 & $-$0.0175 & $-$6.9 ($-$23.7, 6.8) \\
EXT & Stage 1 & Stage+time $\to$ RawCNN-46 & 17 & $-$0.0365 & $-$15.5 ($-$36.9, 0.9) \\
EXT & Stage 1 & SpecLR-46 $\to$ RawCNN-46 & 17 & 0.1437 & 34.6 (23.2, 45.4) \\
\addlinespace[1pt]
EXT & W & Constant $\to$ SpecLR-18 & 17 & 0.0046 & 6.2 ($-$3.0, 17.4) \\
EXT & W & Constant $\to$ SpecLR-46 & 17 & 0.0051 & 6.8 ($-$2.0, 17.8) \\
EXT & W & Constant $\to$ RawCNN-46 & 17 & $-$0.0021 & $-$2.8 ($-$12.1, 4.6) \\
EXT & W & Constant $\to$ Stage-only & 17 & 0.0042 & 5.6 ($-$1.9, 15.4) \\
EXT & W & Constant $\to$ Stage+time & 17 & 0.0042 & 5.6 ($-$0.5, 15.1) \\
EXT & W & Stage-only $\to$ SpecLR-18 & 17 & 0.0004 & 0.6 ($-$3.0, 4.6) \\
EXT & W & Stage+time $\to$ SpecLR-18 & 17 & 0.0004 & 0.6 ($-$4.2, 5.0) \\
EXT & W & Stage-only $\to$ SpecLR-46 & 17 & 0.0009 & 1.3 ($-$1.9, 4.5) \\
EXT & W & Stage+time $\to$ SpecLR-46 & 17 & 0.0009 & 1.3 ($-$3.0, 5.0) \\
EXT & W & Stage-only $\to$ RawCNN-46 & 17 & $-$0.0062 & $-$8.9 ($-$28.9, 2.5) \\
EXT & W & Stage+time $\to$ RawCNN-46 & 17 & $-$0.0063 & $-$8.9 ($-$28.6, 1.8) \\
EXT & W & SpecLR-46 $\to$ RawCNN-46 & 17 & $-$0.0071 & $-$10.3 ($-$31.8, 1.5) \\
\addlinespace[1pt]
EXT & Pre-sleep & Constant $\to$ SpecLR-18 & 17 & 0.0094 & 16.7 (1.7, 39.7) \\
EXT & Pre-sleep & Constant $\to$ SpecLR-46 & 17 & 0.0095 & 16.9 (2.4, 40.9) \\
EXT & Pre-sleep & Constant $\to$ RawCNN-46 & 17 & $-$0.0034 & $-$6.1 ($-$21.3, 6.1) \\
EXT & Pre-sleep & Constant $\to$ Stage-only & 17 & 0.0077 & 13.7 (1.4, 34.9) \\
EXT & Pre-sleep & Constant $\to$ Stage+time & 17 & 0.0077 & 13.8 (2.9, 32.9) \\
EXT & Pre-sleep & Stage-only $\to$ SpecLR-18 & 17 & 0.0017 & 3.5 ($-$2.5, 10.4) \\
EXT & Pre-sleep & Stage+time $\to$ SpecLR-18 & 17 & 0.0017 & 3.4 ($-$4.0, 12.8) \\
EXT & Pre-sleep & Stage-only $\to$ SpecLR-46 & 17 & 0.0018 & 3.7 ($-$1.5, 11.0) \\
EXT & Pre-sleep & Stage+time $\to$ SpecLR-46 & 17 & 0.0018 & 3.7 ($-$2.7, 13.4) \\
EXT & Pre-sleep & Stage-only $\to$ RawCNN-46 & 17 & $-$0.0112 & $-$23.0 ($-$71.8, $-$3.7) \\
EXT & Pre-sleep & Stage+time $\to$ RawCNN-46 & 17 & $-$0.0112 & $-$23.1 ($-$66.8, $-$5.2) \\
EXT & Pre-sleep & SpecLR-46 $\to$ RawCNN-46 & 17 & $-$0.0130 & $-$27.8 ($-$86.4, $-$6.0) \\
\addlinespace[1pt]
EXT & Returning W & Constant $\to$ SpecLR-18 & 13 & $-$0.0237 & $-$13.1 ($-$18.2, $-$5.3) \\
EXT & Returning W & Constant $\to$ SpecLR-46 & 13 & $-$0.0210 & $-$11.6 ($-$18.0, $-$3.3) \\
EXT & Returning W & Constant $\to$ RawCNN-46 & 13 & 0.0061 & 3.4 ($-$19.9, 16.7) \\
EXT & Returning W & Constant $\to$ Stage-only & 13 & $-$0.0167 & $-$9.3 ($-$13.9, $-$2.3) \\
EXT & Returning W & Constant $\to$ Stage+time & 13 & $-$0.0166 & $-$9.2 ($-$13.3, $-$1.0) \\
EXT & Returning W & Stage-only $\to$ SpecLR-18 & 13 & $-$0.0070 & $-$3.5 ($-$6.8, 1.0) \\
EXT & Returning W & Stage+time $\to$ SpecLR-18 & 13 & $-$0.0071 & $-$3.6 ($-$7.4, $-$0.4) \\
EXT & Returning W & Stage-only $\to$ SpecLR-46 & 13 & $-$0.0043 & $-$2.2 ($-$6.8, 4.3) \\
EXT & Returning W & Stage+time $\to$ SpecLR-46 & 13 & $-$0.0043 & $-$2.2 ($-$7.8, 2.8) \\
EXT & Returning W & Stage-only $\to$ RawCNN-46 & 13 & 0.0229 & 11.6 ($-$16.4, 25.6) \\
EXT & Returning W & Stage+time $\to$ RawCNN-46 & 13 & 0.0228 & 11.6 ($-$18.0, 25.9) \\
EXT & Returning W & SpecLR-46 $\to$ RawCNN-46 & 13 & 0.0271 & 13.5 ($-$13.5, 26.8) \\
\bottomrule
\end{longtable}}}

\newcommand{\TableWtwoS}{%
\begin{table}[h]
\centering
\caption{Wake-to-sleep forecasting comparisons (exploratory, development cohort). $R$ (\%) relative to a within-population time reference fitted in the training folds (95\% person-cluster CI). Population sizes (windows/positives): E1 3058/763; E2-pre 3058/132; E2-ret 515/117; E3 1219/582. Feature-based logistic regression and boosting use aperiodic, entropy, DFA and spectral predictors; TCN-30s/TCN-5m: causal temporal CNN on 30\,s or 5\,min of raw EEG; +stage: auxiliary stage head with a state-mixture hazard; Spec-5m: 5\,min log-spectrogram input. The oracle uses scorer-derived time since first sleep and is not a usable forecaster.}
\label{tab:w2s}
\footnotesize
\setlength{\tabcolsep}{3pt}
\begin{tabular}{@{}lllll@{}}
\toprule
Model & E1: W$\to$N1 & E2-pre: W$\to$N2 & E2-ret & E3: N1$\to$N2 \\
\midrule
Feature logistic regression & $-$10.1 ($-$37.7, 13.1) & $-$10.0 ($-$32.7, 5.6) & 8.0 (3.9, 13.1) & 18.7 (4.3, 31.8) \\
Feature boosting & $-$25.0 ($-$66.3, 10.0) & $-$13.4 ($-$43.0, 6.2) & 5.7 ($-$1.5, 13.3) & 9.3 ($-$9.0, 27.5) \\
TCN-30s & $-$1.3 ($-$21.0, 16.3) & $-$40.8 ($-$101.7, $-$3.9) & 15.2 (2.4, 26.3) & $-$3.8 ($-$20.0, 7.8) \\
TCN-5m & 0.1 ($-$18.0, 16.0) & $-$23.5 ($-$63.5, 1.2) & 10.6 ($-$1.9, 21.8) & $-$1.9 ($-$17.7, 10.3) \\
TCN-5m+stage & 0.1 ($-$20.6, 17.7) & $-$18.5 ($-$63.1, 8.3) & 8.3 ($-$7.0, 21.4) & 4.0 ($-$11.5, 16.6) \\
Spec-5m+stage & $-$2.7 ($-$24.7, 12.7) & $-$29.8 ($-$74.2, $-$2.0) & 13.8 (4.3, 24.0) & 20.6 (8.1, 31.2) \\
ORACLE state (diagnostic) & --- & 3.0 (0.0, 8.1) & 12.4 ($-$3.0, 24.9) & 2.1 ($-$1.1, 5.8) \\
\bottomrule
\end{tabular}
\end{table}}

\newcommand{\TablePzeroPone}{%
\begin{table}[h]
\centering
\caption{Wake-to-first-N1 recalibration and personalisation (exploratory, development cohort). Calibrators were fitted by nested person-disjoint cross-fitting (no evaluated person\textquotesingle s labels used by base model, hyperparameters or calibrator). Recalibration variants: intercept-only, intercept-and-slope, and score-plus-elapsed-time. Personalisation variants include a basic temporal encoder, within-person EEG baseline, recurrent history model, calibration-aware loss and their combination. Donor: own history replaced by another person\textquotesingle s. $R$ relative to the within-E1 time reference.}
\label{tab:p0p1}
\footnotesize
\setlength{\tabcolsep}{3pt}
\begin{tabular}{@{}llrrrr@{}}
\toprule
Model & Partition & $R$ (\%) & CIL & Slope & Log loss \\
\midrule
TCN-5m uncalibrated & A & 0.1 ($-$18.0, 16.0) & 0.069 & 0.57 & 0.560 \\
TCN (5 min) + time recalibration & A & 6.2 ($-$2.8, 13.7) & 0.016 & 0.83 & 0.533 \\
TCN-5m+stage uncalibrated & A & 0.1 ($-$20.6, 17.7) & 0.051 & 0.54 & 0.555 \\
TCN + intercept recalibration & A & 3.7 ($-$10.1, 15.5) & $-$0.007 & 0.55 & 0.547 \\
TCN + slope recalibration & A & 6.1 ($-$4.4, 14.2) & 0.003 & 0.83 & 0.532 \\
TCN (5 min, stage head) + time recalibration & A & 7.8 ($-$0.2, 14.6) & $-$0.000 & 0.87 & 0.524 \\
\midrule
Temporal encoder & A & 1.7 ($-$9.8, 10.7) & 0.056 & 0.71 & 0.556 \\
Within-person baseline & A & $-$0.5 ($-$9.5, 7.4) & 0.053 & 0.42 & 0.572 \\
Within-person baseline (donor) & A & 2.5 ($-$2.9, 7.8) & 0.044 & 0.61 & 0.560 \\
Recurrent history & A & 4.6 ($-$3.7, 11.1) & 0.041 & 1.02 & 0.544 \\
Recurrent history (donor) & A & 4.6 ($-$4.0, 11.2) & 0.042 & 1.01 & 0.544 \\
Calibration-aware loss & A & $-$2.0 ($-$10.5, 5.8) & 0.077 & 0.60 & 0.572 \\
Combined model & A & $-$2.9 ($-$10.5, 5.3) & 0.083 & 0.55 & 0.576 \\
Combined model (donor) & A & $-$1.3 ($-$8.1, 6.0) & 0.078 & 0.62 & 0.569 \\
Baseline + time recalibration & A & 4.1 ($-$3.6, 9.9) & $-$0.000 & 0.71 & 0.543 \\
Combined + time recalibration & A & 2.5 ($-$1.2, 6.1) & $-$0.010 & 0.84 & 0.552 \\
\addlinespace[2pt]
Temporal encoder & B & 4.1 ($-$5.3, 12.2) & 0.037 & 0.92 & 0.539 \\
Within-person baseline & B & 2.8 ($-$4.9, 9.5) & 0.020 & 0.79 & 0.545 \\
Within-person baseline (donor) & B & 2.8 ($-$4.8, 9.0) & 0.017 & 0.84 & 0.545 \\
Recurrent history & B & 5.3 ($-$3.7, 13.1) & 0.039 & 1.06 & 0.534 \\
Recurrent history (donor) & B & 5.3 ($-$3.8, 13.2) & 0.039 & 1.05 & 0.534 \\
Calibration-aware loss & B & $-$2.5 ($-$10.8, 5.6) & 0.074 & 0.65 & 0.565 \\
Combined model & B & $-$3.2 ($-$11.3, 4.9) & 0.066 & 0.57 & 0.567 \\
Combined model (donor) & B & $-$2.4 ($-$10.0, 5.2) & 0.065 & 0.61 & 0.563 \\
Baseline + time recalibration & B & 4.8 (0.5, 9.0) & $-$0.012 & 1.05 & 0.534 \\
Combined + time recalibration & B & 0.2 ($-$4.7, 4.6) & $-$0.007 & 0.67 & 0.553 \\
\bottomrule
\end{tabular}
\end{table}}
\long\def\FixTableFourCap#1cohorts , generated#2\FixTableFourEnd#3#4{\def\TableModels{#3\caption{#1cohorts, generated#2}#4}}
\long\def\FixTableFour#1\caption#2#3\FixTableFourEnd{\FixTableFourCap#2\FixTableFourEnd{#1}{#3}}
\expandafter\FixTableFour\TableModels\FixTableFourEnd
\long\def\FixTableTwoCap#1for the second model relative to the first.#2\FixTableTwoEnd#3#4{\def\TableCentral{#3\caption{#1for the first-named forecast or reference relative to the second-named comparator; positive values favour the first-named.#2}#4}}
\long\def\FixTableTwo#1\caption#2#3\FixTableTwoEnd{\FixTableTwoCap#2\FixTableTwoEnd{#1}{#3}}
\expandafter\FixTableTwo\TableCentral\FixTableTwoEnd

\makeatletter
\newcommand{\WideFrozenTable}[1]{{%
  \renewenvironment{table}[1][t]{\@dblfloat{table}[##1]\centering\begin{minipage}{5.5in}}{\end{minipage}\end@dblfloat}%
  #1}}
\makeatother

\icmltitlerunning{Where Does EEG Sleep-Onset Forecasting Skill Arise?}

\begin{document}

\twocolumn[
  \icmltitle{Where Does EEG Sleep-Onset Forecasting Skill Arise?\\ A State-Conditioned Evaluation}
  \begin{icmlauthorlist}
    \icmlauthor{Ali Amini}{ktu}
    \icmlauthor{Mohammad Alijanpour}{ucf}
    \icmlauthor{Kaare B Mikkelsen}{aarhus}
    \icmlauthor{Robertas Damaševičius}{ktu}
  \end{icmlauthorlist}
  \icmlaffiliation{ktu}{Centre of Real Time Computer Systems, Kaunas University of Technology, Kaunas, Lithuania}
  \icmlaffiliation{ucf}{Department of Electrical Engineering and Computer Science, University of Central Florida, Orlando, FL, USA}
  \icmlaffiliation{aarhus}{Department of Electrical and Computer Engineering, Aarhus University, Aarhus N 8200, Denmark}
  \icmlcorrespondingauthor{Ali Amini}{\mbox{ali.amini@ktu.edu}}
  \icmlkeywords{evaluation, proper scoring rules, event forecasting, EEG, sleep onset}
  \vskip 0.3in
]
\printAffiliationsAndNotice{}

\begin{abstract}
Sleep is scored afterwards in 30-second epochs as wake (W), the transitional stage N1, or N2. We study EEG forecasts, issued every 30\,s from the preceding 30\,s of EEG alone, of whether the first scored N2 epoch will begin within five minutes, on Sleep-EDF. A Brier score pooled over decisions made before and after the input is scored N1 does not show when the gain arose. In our own prespecified one-shot held-out comparisons, the spectral model SpecLR-18 improved on a training-prevalence constant by 18.3\% in the internal test, whereas its prespecified external result was inconclusive. Joining the retrospectively scored stage of each input epoch to the same frozen predictions, almost all of that internal-test Brier difference was attributed to N1-scored decisions; a diagnostic reference told only the scored stage reached 26.3\% in the internal test and 29.5\% in the external cohort, above the spectral forecasts in point estimate though not every difference excluded zero; and within-stage EEG advantages seen in single cohorts were not consistent across cohorts, with no consistent wake-stratum improvement. These references use labels unavailable online and cannot reveal what a model represents; they show why pooled forecasting scores need state-conditioned interpretation before being read as early anticipation.
\end{abstract}

\section{Introduction}
\label{sec:intro}

In polysomnography, a scorer labels each 30-second epoch after recording as wake (W), transitional N1, N2 or a deeper sleep stage. Here an EEG forecaster, without access to those labels, uses the preceding 30\,s of EEG every 30\,s to predict whether the \emph{first scored N2 epoch} will begin within five minutes. This is distinct from forecasting first N1 or issuing a warning only while the person is still awake. The distinction creates an evaluation question: decisions made before first N2 include both W and N1 inputs, so improvement averaged over them need not imply earlier anticipation during wakefulness.

The five-minute event rate is several times higher on inputs already scored N1 than on inputs scored W, and most positive pre-N2 decisions occur in N1. A proper score such as the Brier score \citep{brier1950,gneiting2007scoring}, pooled over all decisions, validly measures accuracy on this defined task. But it cannot say \emph{when} the gain arose: changing the forecast probability after entry into N1 can improve the score without establishing useful anticipation from wake. Being more accurate is not the same as being earlier.

This paper reports our own prespecified, one-shot held-out evaluations of three EEG forecasters on Sleep-EDF and then re-examines the same recorded predictions. Two spectral logistic regressions and a generic convolutional network, frozen before evaluation, were compared under prespecified rules with a training-prevalence constant and sequentially with one another, on held-out persons from the same study and on a second study in the same dataset family (\emph{Q1}). Joining the scorer's stage of each input epoch to those predictions, with no model rerun, we then ask where across W- and N1-scored inputs the pooled improvement arose (\emph{Q2a}), how references using the scored stage, with or without elapsed time, fare on the same decisions (\emph{Q2b}), and whether any EEG skill beyond stage and elapsed time was consistent across cohorts within a stage (\emph{Q3}); these analyses were formulated after the Q1 results had been examined and are retrospective and descriptive (Fig.~\ref{fig:task}b).

Most of the internal-test improvement was earned on N1-scored inputs; a reference told only the scored stage matched or exceeded the spectral forecasts in point estimate in the internal test and the external cohort; within-stage advantages found in one cohort were not consistent across cohorts; and no model consistently improved on the stage references while the person was still scored awake. The contribution is evaluative, built from established scoring and multistate ingredients \citep{murphy1973partition,graf1999assessment,putter2007multistate,vancalster2019calibration}: a state-conditioned reading of a pooled event-forecast score, exact and non-causal in its attribution, demonstrated on one case. It proposes no architecture and reproduces no outside study; it shows neither that EEG forecasting fails nor that these models merely recognize sleep stage, but what a pooled score does and does not license.

\section{Related work}
\label{sec:related}

\textbf{Sleep-onset forecasting.} \citet{li2025falling} describe the EEG wake-to-sleep transition using a bifurcation framework and demonstrate real-time prediction of sleep progression, defining sleep onset by sustained N2; their prediction target and evaluation differ from our first-epoch five-minute event risk. \citet{su2026presleep} predict NREM2 within 5\,min from alpha--theta dynamics and report classification accuracy. \citet{beppu2025latency} relate wake-EEG indices to N1 and N2 latencies in daytime naps; MEDi-SOL \citep{oh2024medisol} predicts the distribution of sleep-onset latency; SSF-SET \citep{kweon2026ssfset} forecasts stage sequences on Sleep-EDF. Sleep foundation models \citep{huang2026sleepgpt,thapa2026sleepfm} learn general-purpose representations and were not used here. These works target sustained N2, first N2, latencies or stage sequences and report pooled accuracy; our study instead asks which scored input stages contribute to the pooled skill of one first-N2 forecast, and how references told the retrospectively scored stage change its interpretation.

\textbf{Landmark prediction and multistate models.} Predicting an event at successive landmarks, conditional on being event-free at each, is the landmarking approach of \citet{vanhouwelingen2007landmark}; \citet{graf1999assessment} define the corresponding time-dependent Brier score. Sleep onset is naturally a multistate process (W, N1, N2, with returns to wake) in which transition probabilities depend on the current state \citep{putter2007multistate}. The stage references used here (\S\ref{sec:formulation}) are the simplest state-conditional landmark models of this kind.

\textbf{Scoring rules, calibration and forecast evaluation.} We rely on the Brier score and its decomposition \citep{brier1950,murphy1973partition}, the theory of proper scoring rules \citep{gneiting2007scoring}, calibration as a separate performance dimension \citep{vancalster2019calibration,collins2024tripodai} and the observation that calibration degrades under dataset shift \citep{ovadia2019trust}. Seizure forecasting evaluates probabilistic EEG forecasts against chance or surrogate comparators \citep{winterhalder2003spc,proix2021seizure}; \citet{winterhalder2003spc} require sensitivity, false-prediction rate and horizon to be reported together. Pooling out-of-fold scores across cross-validation folds can bias AUROC when fold-specific predictions differ in level \citep{forman2010apples}; our exploratory analysis encountered this issue (App.~\ref{app:w2s}). App.~\ref{app:priorart} maps each component of our analysis to prior work.

\section{Problem formulation and evaluation}
\label{sec:formulation}
\label{sec:estimands}

\begin{figure*}[t]
\centering
\includegraphics[width=5.5in]{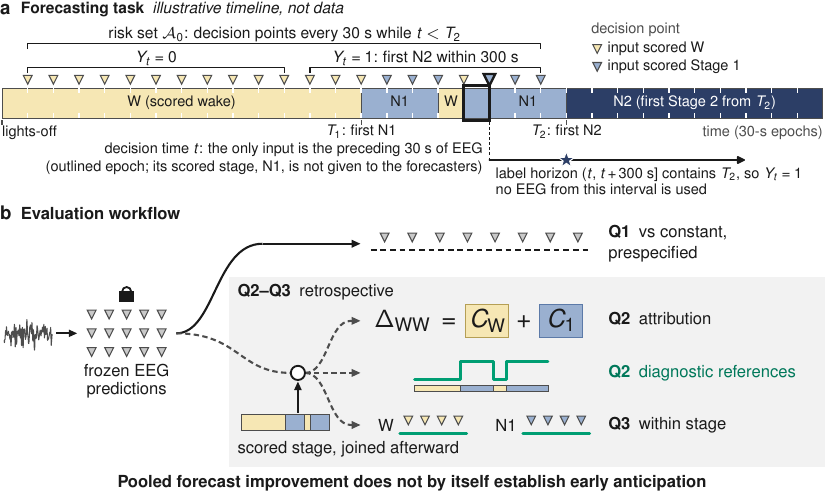}
\caption{Forecasting task and evaluation workflow. (a) Illustrative timeline, not data. Decision points (triangles) occur every 30\,s from lights-off until the first scored N2 epoch at $T_2$ and form the risk set $\mathcal{A}_0$. A decision at time $t$ uses only the preceding 30\,s of EEG (outlined epoch) and is positive ($Y_t=1$) if $T_2$ falls within the horizon $(t,t+300\,\mathrm{s}]$. Each triangle takes the colour of the scored stage of its input epoch (W, including wake recurring after $T_1$, or N1, labelled Stage~1 in the legend); this stage is never given to the EEG forecasters but defines the post-hoc strata. (b) Schematic, not data: waveform, triangles, step levels and coloured chips are not numerical fractions. Held-out predictions of frozen TRAIN-fitted models, from EEG alone, were recorded once (padlock; SpecLR-46 and RawCNN-46 have none on VAL). Solid: prespecified Q1 comparisons with the training-prevalence constant. Grey: retrospective, descriptive analyses (Q2, Q3) reusing the same predictions with no model rerun; the scored stage (W or N1) is joined only afterward (open circle) and never supplied to EEG forecasters. Q2: $\dww=C_{\mathrm{W}}+C_1$ exactly divides the pooled improvement over the constant (accounting, not causal explanation) and compares EEG with the diagnostic TRAIN-fitted stage references, Stage-only (drawn) and Stage+time (not drawn), which are neither deployable comparators nor performance ceilings; Q3 repeats the comparison within each stage. Secondary outcomes (\S\ref{sec:resE}) and TRAIN-only exploratory analyses (App.~\ref{app:w2s}) are not drawn.}
\label{fig:task}
\end{figure*}

\textbf{Task and risk set.} For a night with lights-off at time $0$, decision points are $t\in\{30,60,\dots\}$\,s, and each uses only the EEG window $[t-30,t)$ (Fpz-Cz and Pz-Oz, 100\,Hz). Let $T_2$ be the onset of the first scored N2 epoch, $T_1$ the onset of the first scored sleep epoch (always N1 in these data), and $S(t)\in\{\mathrm{W},1\}$ the scored stage of the input epoch. For night $r$ of person $i$, the primary prediction target has risk set $\mathcal{A}_{0,ir}=\{(i,r,t):t<T_{2,ir}\}$ and label $Y_{ir,t}=\mathbf{1}\{T_{2,ir}\le t+h\}$ with $h=300$\,s (Fig.~\ref{fig:task}); $\mathcal{A}_0$ pools the analysed eligible nights. An EEG forecaster's output $p_{ir,t}$ estimates $P(T_{2,ir}\le t+h\mid T_{2,ir}>t,\ \text{EEG on }[t-30,t))$; scored stages, $T_1$, $T_2$ and eligibility are retrospective and define only labels, risk sets and strata (App.~\ref{app:defs}). Whether N2 can be anticipated while the input is still scored W concerns the W-scored decisions of this same risk set; forecasting the first N1 epoch is a different target, examined only in a TRAIN-only exploratory analysis (App.~\ref{app:w2s}; estimands in App.~\ref{app:defs}).

\textbf{Scores.} For a reference $p^{\mathrm{ref}}$ and a candidate $p^{\mathrm{c}}$ on a set $\mathcal{A}$ of decisions $(i,r,t)$, the \emph{pooled} (window-weighted) Brier improvement averages the loss difference over every decision in $\mathcal{A}$, W- and N1-scored alike, with $\ell^{m}_{ir,t}=(p^{m}_{ir,t}-Y_{ir,t})^2$,
\begin{equation}\label{eq:pooled_improvement}
\dww=\frac{1}{|\mathcal{A}|}\sum_{(i,r,t)\in\mathcal{A}}\big(\ell^{\mathrm{ref}}_{ir,t}-\ell^{\mathrm{c}}_{ir,t}\big),
\end{equation}
and the Brier skill is $\R=\dww/\bar B_{\mathrm{ref}}$, with $\bar B_{\mathrm{ref}}$ the mean Brier score of the reference. Every reference forecast lies strictly in $(0,1)$, so $\bar B_{\mathrm{ref}}>0$; positive values favour the candidate. We report $\dww$ beside $\R$ because equal skill corresponds to different absolute changes when $\bar B_{\mathrm{ref}}$ differs, and skills relative to different references cannot be subtracted, whereas $\dww$ values on identical decisions can. $\dse$ averages over persons each person's mean difference, pooling that person's analysed nights (App.~\ref{app:defs}).

\textbf{Original reference.} Each prespecified evaluation compared against a \emph{constant} equal to the training prevalence. It was chosen during development because no elapsed-time-only reference beat it in cross-validation; it was not intended as the strongest possible reference.

\textbf{Stage attribution.} For $s\in\{\mathrm{W},1\}$, the stage contribution $C_s$ restricts the sum in Eq.~\eqref{eq:pooled_improvement} to decisions with $S(t)=s$, retaining the pooled denominator $|\mathcal{A}_0|$. Hence $\dww=C_{\mathrm{W}}+C_1$ exactly. The identity locates the pooled difference; it does not explain it. Because the constant sits at the pooled prevalence, nearer the W event rate than the N1 rate, even a reference that only reports each stage's event rate would take most of its improvement over the constant from N1-scored decisions; a large $C_1$ is therefore expected for any model whose outputs rise on N1-like EEG, and shows neither that the model detects N1 nor that N1 causes the gain. The share observed for a model is measured, not implied by this arithmetic. Within-cell skill $\R_s$ uses the same reference restricted to the cell.

\textbf{Stage references.} A reference that is told the scored stage asks how much of a pooled improvement is attainable from current-stage information alone. A stage-informed reference beat the \mbox{constant} in development cross-validation; elapsed-time-only references did not (App.~\ref{app:secondary_intervals}). We fitted two such references on TRAIN only and applied them unchanged to the held-out predictions. \emph{Stage-only} assigns the Laplace-smoothed TRAIN event rate of each scored input stage $s$, $(k_s+1)/(n_s+2)$: 0.070 for W inputs and 0.477 for N1 inputs. \emph{Stage+time} is an L2-penalised logistic regression on the N1 indicator and a cubic B-spline of log elapsed time (5 knots, $C=10$, no interaction), selected from 12 configurations by person-disjoint inner cross-validation; elapsed times outside the TRAIN support (30 to 6{,}990\,s) are clamped to its boundary. Both use the retrospectively scored stage, unavailable to an online forecaster, so they are diagnostics, not deployable comparators; and a forecaster may exploit information they ignore, so they are not upper bounds.

\textbf{Uncertainty and discrimination.} Intervals are 95\% percentile bootstraps over persons \citep{davison1997bootstrap}; prespecified results keep their original bootstrap, retrospective contrasts use new seeds (10{,}000 replicates; App.~\ref{app:stat_supplement}), and tables state which is shown (the two can differ in the last digit). Discrimination is reported as within-recording AUROC (prespecified) and as cross-person AUROC matched on elapsed time, within prespecified coarse bins in the original held-out evaluation (Table~4) and at the same exact 30\,s landmark in the retrospective analysis; these are different estimators (App.~\ref{app:defs}). Any increasing function of elapsed time attains a high within-recording AUROC, and within a scored stage the stage references are constant at each exact landmark, so their matched AUROC is 0.5 by construction.

\section{Experimental setting}
\label{sec:data}
\WideFrozenTable{\TableFrozen}

\textbf{Data and cohorts.} Table~\ref{tab:frozen} summarizes the original Q1 comparisons; the following cohort and model definitions explain its labels. Sleep-EDF Expanded v1.0.0 \citep{kemp2000sleepedf,pollard2026physionet} contains the Sleep Cassette study (SC; home recordings of healthy adults) and the Sleep Telemetry study (ST; hospital recordings on placebo and temazepam nights), scored with R\&K rules \citep{rechtschaffen1968manual}; we write N1 and N2 for R\&K Stages 1 and 2, the names kept in the tables and figure legends. A night is eligible if lights-off lies within the signal, no scoring gap precedes first N2, N2 is not active at lights-off and the stage at lights-off is W (flow in App.~\ref{app:cohort}, Table~\ref{tab:cohort}). SC persons were split once, by salted hash, into three person-disjoint groups: TRAIN (84 nights, 46 persons) for development; VAL (15 persons), the first one-shot gate for the pilot spectral model only, earlier used to evaluate non-EEG references; and TEST (15 persons), the internal one-shot test for all three EEG models. EXT comprises all analysable ST nights (28 nights, 17 persons), a different study, setting and intervention within the same dataset family, so it probes cross-study transfer rather than an independent dataset. VAL and TEST primary analyses used one predeclared night per person; all-nights analyses were predeclared sensitivity analyses. One VAL night reached first N2 under 330\,s after lights-off, so it had seven rather than ten positive decisions (VAL: 147; App.~\ref{app:defs}).

\textbf{Models and their roles.} The three EEG forecasters are evaluation instruments, not proposed architectures; each was frozen before its held-out assessment (details in App.~\ref{app:defs}). \emph{SpecLR-18} is the prespecified spectral logistic-regression pilot, using ten band-power and amplitude features from two channels and one night per 18 TRAIN persons (1{,}271 windows); it alone was evaluated on VAL and was carried unchanged into TEST and EXT. \emph{SpecLR-46} applies the same spectral recipe to all 84 nights of the 46 TRAIN persons, separating training-roster size from feature family. \emph{RawCNN-46} is a generic multiscale 1D network on the raw 30\,s waveform (114{,}817 parameters; three-seed mean), trained on the same 46-person roster. Neither comparison isolates an architecture effect: SpecLR-18--RawCNN-46 confounds architecture with roster, and SpecLR-46--RawCNN-46 holds the roster fixed but not representation, effective spectral coverage, model class, capacity or training (App.~\ref{app:defs}). Six SpecLR-18 pilot participants were selected by observed sleep latency, so that sample is outcome-dependent rather than random. The CNN replaced a predeclared task-specific network after development cross-validation (App.~\ref{app:ablation}). \emph{Constant} is the training-prevalence reference. Stage-only and Stage+time were fitted on all 46 TRAIN persons and, as a sensitivity check, also refitted on the 18-person roster.

\textbf{Protocol and inferential status.} The first-N2 endpoint was inherited from an earlier task on a different public benchmark without N1 annotations and retained as the fixed target on Sleep-EDF; it was not chosen because first N2 marks stable sleep or the first loss of wakefulness. Model parameters, participant rosters, comparison rules and analysis code were fixed and documented before the designated held-out evaluations (App.~\ref{app:protocol}). The primary comparison was SpecLR-18 against the training-prevalence constant; the CNN comparison against SpecLR-18 was evaluated sequentially only when the primary comparison met its prespecified criteria, and the equal-roster contrast against SpecLR-46 was descriptive. Eligibility and event timing had been determined from scored annotations before model assessment, so the held-out populations were model-naive but not fully outcome-blind. No held-out model comparison was rerun. After the initial held-out results had been examined, we specified the stage-conditioned analyses and applied them to the recorded predictions without refitting: the current-stage composition of the risk set, the attribution of the original score differences to scored input stages, and the stage references fitted on development participants. Everything from \S\ref{sec:resB} onward is therefore retrospective and descriptive, not an independent prospective test of the new research questions.

\section{Results}
\label{sec:results}

\subsection{Q1: Prespecified EEG comparisons established improvement over a constant}
\label{sec:resA}

\begin{figure*}[t]
\centering
\includegraphics[width=5.5in]{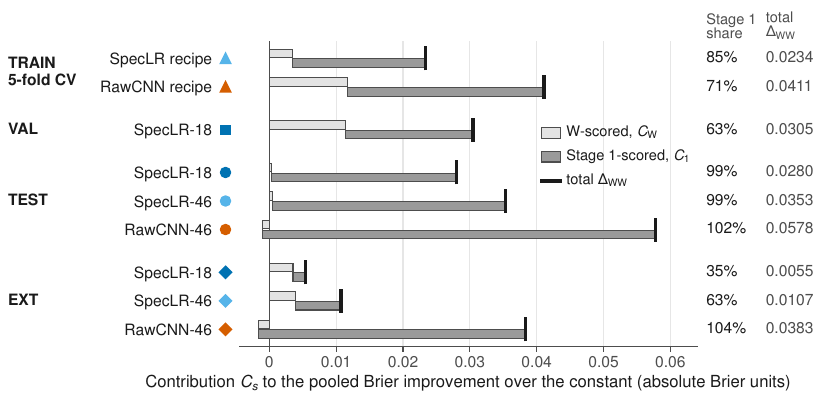}
\caption{Exact attribution of the pooled Brier improvement over the constant to the scored stage of the input epoch, in absolute Brier units. In each row the light bar runs from 0 to the wake contribution $C_{\mathrm{W}}$ (negative for RawCNN-46 in TEST and EXT) and the dark bar from $C_{\mathrm{W}}$ to $C_{\mathrm{W}}+C_1=\dww$ (vertical tick); the Stage~1 share $C_1/\dww$ can therefore exceed 100\%. TRAIN: 5-fold cross-validated recipes. VAL, TEST, EXT: original recorded predictions; SpecLR-46 and RawCNN-46 were not scored on VAL. Marker shape denotes the cohort. Point values; stratum-specific skill with intervals is in Table~\ref{tab:r2full}. The attribution is an accounting identity: it locates the difference; it does not explain it.}
\label{fig:attr}
\end{figure*}

The prespecified tests asked whether EEG forecasts improved on the training-prevalence constant, not whether they exceeded information associated with the scored stage. SpecLR-18 met its prespecified primary criteria, which include a donor-EEG negative control (App.~\ref{app:protocol}), in VAL (22.3\% Brier skill) and TEST (18.3\%, 95\% person-cluster interval 8.8--28.8\%; Table~\ref{tab:frozen}). In TEST, RawCNN-46 also met the sequential secondary criterion against SpecLR-18 (23.8\%, interval 4.7--41.2\%). However, this comparison changes both model and training roster; with the 46-person roster held fixed, the point advantage over SpecLR-46 (19.1\%) had an interval including zero, so a separate architectural advantage was not established in TEST.

In EXT, the prespecified primary comparison of \mbox{SpecLR-18} was inconclusive (3.4\%, interval $-$0.9 to 7.6\%), so the conditional CNN comparison was not formally tested in EXT; descriptively, RawCNN-46 exceeded SpecLR-46 there by 18.5\% (interval 8.2--26.7\%), an equal-roster contrast that was never a prespecified test. SpecLR-18 retained a within-recording AUROC near its TEST value while underpredicting the external event rate (\S\ref{sec:resE}). Q1 thus establishes improvement over the constant on the pooled first-N2 task; it does not say on which inputs that improvement was earned.

\subsection{Q2a: Most internal-test gains accrued on N1-scored inputs}
\label{sec:resB}

\begin{figure*}[t]
\centering
\includegraphics[width=\textwidth]{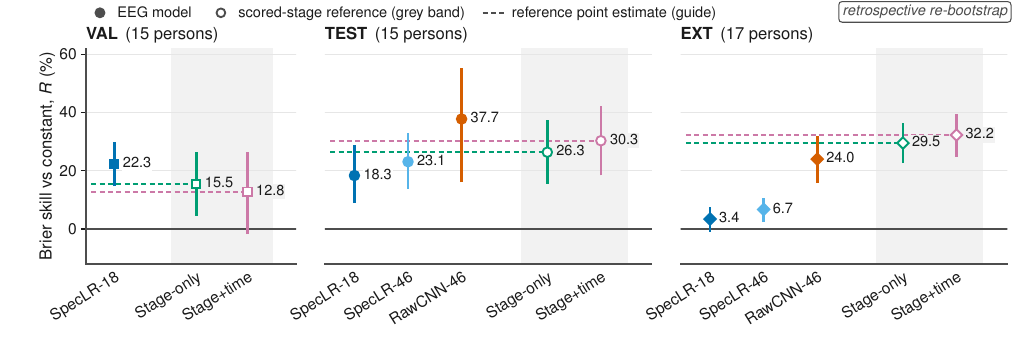}
\caption{Brier skill relative to the constant, $R$ (\%), for EEG forecasts (filled markers) and development-fitted scored-stage references (open markers on the grey band), one panel per cohort on a shared axis, with 95\% participant-cluster intervals from the retrospective re-bootstrap of the original predictions (Table~\ref{tab:central}; the original prespecified intervals are in Table~\ref{tab:frozen}). Dashed guides carry each reference's point estimate across the EEG columns; printed numbers are point estimates. The references use the retrospectively scored stage and are diagnostic, not deployable. Marker shape denotes the cohort. VAL evaluated SpecLR-18 only, so SpecLR-46 and RawCNN-46 have no VAL columns (not evaluated, not zero).}
\label{fig:refs}
\end{figure*}

\WideFrozenTable{\TableCentral}

If the pooled improvement had been earned while the person was still awake, it would appear on W-scored inputs. Among positive decision points, however, the input epoch was already scored N1 in 62\%, 71\% and 70\% of VAL, TEST and EXT windows, respectively, and the event rate on N1 inputs was 43--59\%, compared with 7--8\% on wake inputs. Figure~\ref{fig:attr} partitions the \emph{original} pooled Brier difference by the scored input stage (TRAIN rows: development cross-validation of the same recipes); intervals are in App.~\ref{app:state_strata}.

\textbf{TEST.} N1 inputs contributed 99\% of each spectral model's improvement (SpecLR-18: 95\% interval 81--129\%) and 102\% of RawCNN-46's (above 100\% because the wake contribution, whose interval includes zero, was slightly negative). No EEG model established improvement over the constant on wake-scored inputs. \textbf{EXT.} The CNN likewise derived 104\% (93--120\%) of its pooled gain from N1 inputs and SpecLR-46 63\% (36--124\%); SpecLR-18's pooled gain was too small for a stable share. \textbf{VAL.} N1 inputs supplied 63\% (38--89\%) of SpecLR-18's improvement and wake inputs the rest. Absolute contributions with intervals are in App.~\ref{app:stat_supplement}. The stage concentration is therefore an empirical model-and-cohort pattern rather than a universal fact about EEG forecasts.

A wake-only improvement for the CNN recipe was suggested in development cross-validation but not established for the frozen CNN in either held-out TEST or EXT. These shares locate the score difference (\S\ref{sec:formulation}); they do not show whether the EEG carried information beyond the scored stage itself.

\subsection{Q2b: Stage-only and Stage+time references change the comparison}
\label{sec:resC}

\begin{figure*}[t]
\centering
\includegraphics[width=\textwidth]{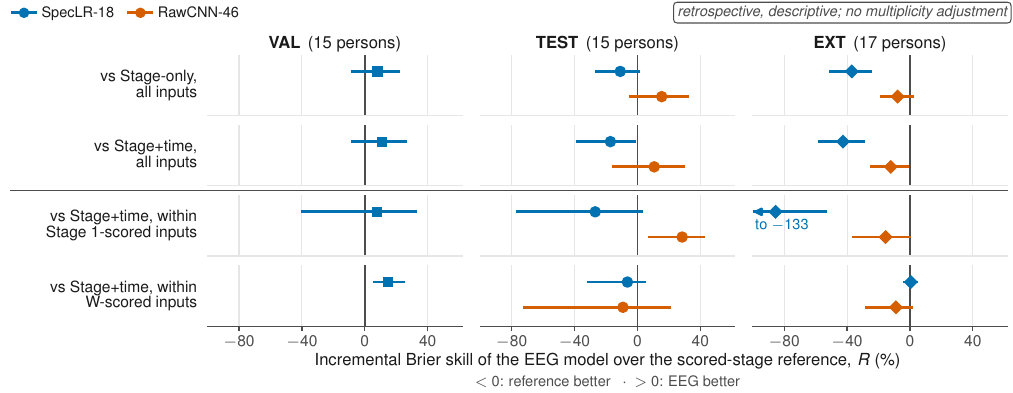}
\caption{Incremental Brier skill of EEG forecasts relative to the scored-stage references, arranged to compare cross-cohort patterns along each row: rows are contrasts (top two: all inputs; bottom two: within the scored input state, against Stage+time), columns are cohorts, and each cell shows SpecLR-18 (upper) and RawCNN-46 (lower) with 95\% participant-cluster re-bootstrap intervals (Table~\ref{tab:central}, last four rows). Positive values favour the EEG model; marker shape denotes the cohort. All cells share the axis range $-100$ to 62\%; the only interval extending beyond it (EXT, SpecLR-18, Stage~1 inputs; lower bound $-$133\%) is truncated at $-100\%$ and marked with an arrow. RawCNN-46 was not evaluated on VAL, so VAL cells show SpecLR-18 only (not evaluated, not zero). Within W-scored inputs, the only interval excluding zero is SpecLR-18 on VAL (14.6\%, 5.2 to 25.8\%), which was not observed in TEST or EXT (between-cohort contrasts in App.~\ref{app:stat_supplement}). These retrospective contrasts are descriptive and not adjusted for multiplicity.}
\label{fig:incr}
\end{figure*}

If information in the current scored stage explained the pooled gain, a reference given that stage should perform comparably. Both stage references were fitted on TRAIN and transferred unchanged to the held-out cohorts (Fig.~\ref{fig:refs}; Table~\ref{tab:central}). In Table~\ref{tab:central}, the pooled rows concern Q2b; the final N1- and W-specific rows anticipate the within-stage question Q3.

\textbf{TEST.} Stage-only and Stage+time improved on the constant by 26.3\% and 30.3\%, respectively, above both spectral models' point estimates; against SpecLR-18 the Stage+time difference excluded zero ($-$17.1\%, $-$39.3 to $-$0.6\%), the Stage-only difference did not. RawCNN-46 had a larger point estimate than either reference, with intervals including zero. \textbf{EXT.} The stage references reached 29.5\% and 32.2\%, above all three EEG models, and each EEG-to-Stage+time difference excluded zero on the negative side. In \textbf{VAL} the references' point estimates were lower (15.5\% and 12.8\%) and the spectral model's point advantage over them had intervals including zero; with 15 persons per cohort the intervals are wide. A sensitivity refit of Stage-only on the 18-person development roster did not reverse the TEST/EXT pattern (App.~\ref{app:stage_reference}).

Falling below a reference that is told the stage does not identify the model's internal representation; it shows that improving on a constant is insufficient evidence of skill beyond information associated with the scored input stage. Whether any such skill exists must be sought within a stage.

\subsection{Q3: Within-stage EEG advantages were not consistent across cohorts}
\label{sec:resD}

Any EEG information beyond the current stage must show within a stage, where Stage-only reduces to a constant and Stage+time to a function of elapsed time. We therefore examined incremental EEG skill within each scored input stage against these references (Fig.~\ref{fig:incr}; Table~\ref{tab:central}, last two rows; full person-cluster intervals in Table~\ref{tab:r3bfull}).

\textbf{N1.} In TEST, RawCNN-46 improved on Stage+time by 28.5\% (95\% interval 7.0--43.3\%), but not in EXT ($-$15.5\%, interval $-$36.9 to 0.9\%); in a retrospective direct comparison, the TEST$-$EXT difference in absolute Brier improvement was 0.101 (95\% interval 0.037--0.164; Bonferroni-adjusted for three contrasts 0.023--0.178). The spectral model did not establish a within-N1 advantage in any held-out cohort. \textbf{Wake.} SpecLR-18 exceeded Stage+time in VAL (14.6\%, interval 5.2--25.8\%) but not in TEST or EXT; the VAL$-$EXT difference was 0.010 (0.002--0.036; adjusted 0.0004--0.043, borderline) and the VAL$-$TEST difference 0.015 (0.002--0.041; adjusted interval including zero). RawCNN-46 did not establish a wake-only improvement in either held-out cohort. These contrasts were chosen after the per-cohort results had been seen, with the procedure fixed before computation (App.~\ref{app:stat_supplement}); because the cohorts also differ in setting, drug condition, prevalence and selection, they describe heterogeneity of recorded improvement, not a mechanism or independently confirmed hypotheses.

Exact-landmark-matched cross-person AUROC on wake inputs nevertheless exceeded chance for RawCNN-46 in TEST and SpecLR-18 in VAL. That ranking result does not entail lower Brier risk against Stage+time. N1 AUROC for the CNN in TEST involved only six people contributing negatives and was not assigned a confidence interval.

\subsection{Secondary outcomes: calibration, ranking and warnings disagree}
\label{sec:resE}

\textbf{Probability level and ranking.} In EXT, SpecLR-18 retained within-recording AUROC near its TEST value but predicted probabilities averaging 0.130 below the observed event rate, with an unfavorable between-person contribution to the Brier-score difference (App.~\ref{app:secondary_intervals}). This is compatible with probability-level degradation under shift but does not identify its cause \citep{ovadia2019trust}. The constant was itself poorly calibrated within N1, so correcting forecast probability levels can contribute to a model's N1 Brier gain.

\textbf{Separate TRAIN-only exploration.} Forecasting first N1 from wake EEG was explored as a different target using development-cohort cross-validation, not as a held-out first-N2 test (App.~\ref{app:w2s}). It did not establish reliable Brier improvement over a time-only reference; the recalibration and warning-threshold tradeoffs are documented in Apps.~\ref{app:w2s} and \ref{app:secondary_intervals} \citep[cf.][]{winterhalder2003spc}.

\section{Discussion}
\label{sec:discussion}

Where did the skill arise? In the internal test, almost entirely on inputs already scored N1; in the external cohort, almost entirely so for the CNN and less clearly for the spectral models; in the validation cohort, partly on wake inputs. The pooled Brier skill on the first-N2 task remains a valid answer to the prespecified question; what it cannot show by itself is that EEG provided useful warning while the person was still awake, because the score averages decisions across stages whose event rates differ several-fold, and an accounting attribution locates the gain without explaining it.

The stage references make this limit concrete: in TEST and EXT a reference told only the scored stage matched or exceeded the spectral models' pooled improvement in point estimate. Two explanations are compatible: the models may exploit EEG features that track the scored stage, or information that varies within a stage or with proximity to the event. The within-stage results bear on both without deciding: RawCNN-46 exceeded Stage+time within N1 in TEST but not in EXT, a retrospective cohort difference that cannot isolate setting, drug condition or selection, and no model consistently exceeded the references on wake inputs. The references are not online systems and not ceilings; they expose the gap between improvement over the original constant and improvement over current-stage information, and establishing learned stage recognition would need evidence they do not provide.

Ranking, probabilistic calibration and alarm utility also disagreed on the same prediction sequences. A more informative evaluation should therefore report risk-set composition, attribution of pooled score differences to relevant states, within-state comparisons against clearly specified references, and distinct calibration, discrimination and warning outcomes, with their uncertainty and the timing of the research questions stated next to the findings.

These recommendations rest on one Sleep-EDF case study and do not cover seizure forecasting or acute deterioration. Broader validation would require independent datasets or prospective tests of the state-conditioned questions.

\section{Limitations}
\label{sec:limitations}

\textbf{Population and task.} The analysis covers one Sleep-EDF dataset family, two EEG channels, R\&K scoring and a five-minute first-N2 horizon. The held-out cohorts include only 15--17 persons each, the independent units: decisions within a night are serially dependent, EXT includes two nights for 11 persons, and N1-input contrasts rest on 13 (VAL), 14 (TEST) and 17 (EXT) persons; within-wake analyses sometimes have 44--84 positive windows. Wide person-cluster intervals are approximate with so few clusters and do not establish equivalence or absence of information, and these results need not generalize to other sensors, settings, populations or event targets. Every night began at an intended lights-off in bed and reached first N2; driving or shift-work drowsiness, real-time acquisition and latency, alarm thresholds and clinical use were not evaluated, so no operational warning performance is established.

\textbf{Timing and selection.} The state-conditioned scientific questions were developed after inspecting earlier held-out outcomes. Many strata and model contrasts were inspected without multiplicity adjustment, so their intervals are descriptive; the three direct between-cohort contrasts were selected after earlier results had been seen, so even their Bonferroni-adjusted intervals do not make them confirmatory. Eligibility and event times relied on scored outcomes, every included night reached N2, and VAL had earlier been used to evaluate non-EEG references.

\textbf{Confounding and mechanisms.} The SpecLR-18 roster is outcome-selected and smaller than the development set of SpecLR-46 and RawCNN-46, so the unequal-roster comparison does not identify an architecture effect. The CNN also replaced an earlier design after development cross-validation; repeated partitions of the same development participants are not independent tests. Spectral sensitivity and perturbation results do not establish a physiological mechanism or network representation.

\textbf{Reproducibility.} The paper-specific implementation is maintained in a separate MIT-licensed code repository, as indicated in App.~\ref{app:code_access}. Synthetic tests and TRAIN-only equivalence checks have passed; independent clean-room verification is still pending. Person-cluster bootstrap intervals describe sampling variability under the recorded analysis; they do not remove retrospective hypothesis-selection risk.

\section{Conclusion}
\label{sec:conclusion}

For first-N2 forecasting in these Sleep-EDF cohorts, most of the internal-test Brier improvement over a constant arose on inputs already scored N1; a reference told only the scored stage matched or exceeded the spectral forecasts in point estimate in the internal test and the external cohort; and no model consistently improved on that reference while the person was still scored awake. These results are consistent with much of the pooled skill reflecting the current scored stage, do not establish anticipation of sleep from wake, and neither prove nor exclude stage recognition. A pooled score alone does not say \emph{when} predictive information becomes available. State-conditioned diagnostics make that limit of interpretation visible.

\section*{Impact Statement}
Sleep-onset forecasts are discussed for drowsiness monitoring and sleep-health applications. Our results caution that much of a pooled Brier-score improvement over a constant can accrue after the input has already been scored in an intermediate sleep state, without establishing what the model internally represents; good ranking may also coexist with poorly calibrated probabilities in a new population. Over-reliance on such forecasts in safety-critical settings could cause harm. The work uses public, de-identified data under its licence and involves no new human-subjects research.



\newpage
\appendix
\onecolumn
\raggedbottom

\section{Cohort flow}
\label{app:cohort}
\TableCohort

\section{Prespecified comparisons and protocol deviations}
\label{app:protocol}
\textbf{Prespecified decision criteria.} The primary comparison met its success criteria when the lower 95\% bound of $\dww$ exceeded zero, $\R\ge5\%$, the person-equal difference $\dse>0$, the lower within-recording AUROC bound exceeded 0.5, and the donor-EEG control had a non-positive lower interval bound. The prespecified failure criteria were an upper bound on $\R$ below 5\% or non-positive values for both point differences; otherwise the result was indeterminate. Failed permutation or integrity checks invalidated the evaluation. The CNN comparison was evaluated only if the primary criteria were met, within each cohort's predefined sequence; the matched-training-population contrast and all state-specific analyses were descriptive. \textbf{Protocol deviations.} The CNN output bias was initialized at the training prevalence, and RawCNN replaced the originally planned temporal architecture after development cross-validation (App.~\ref{app:ablation}). \textbf{Retrospective analyses.} The stage-conditioned analyses were motivated by subsequent reviews of earlier manuscript drafts. Their analysis plans and computational checks were completed before re-examining the held-out predictions; these steps do not make the hypotheses prospective. Detailed execution records, including the original protocol identifiers, are retained in the reproducibility archive.\par

\TableModels

\begin{figure}[t]
\centering
\includegraphics[width=\textwidth]{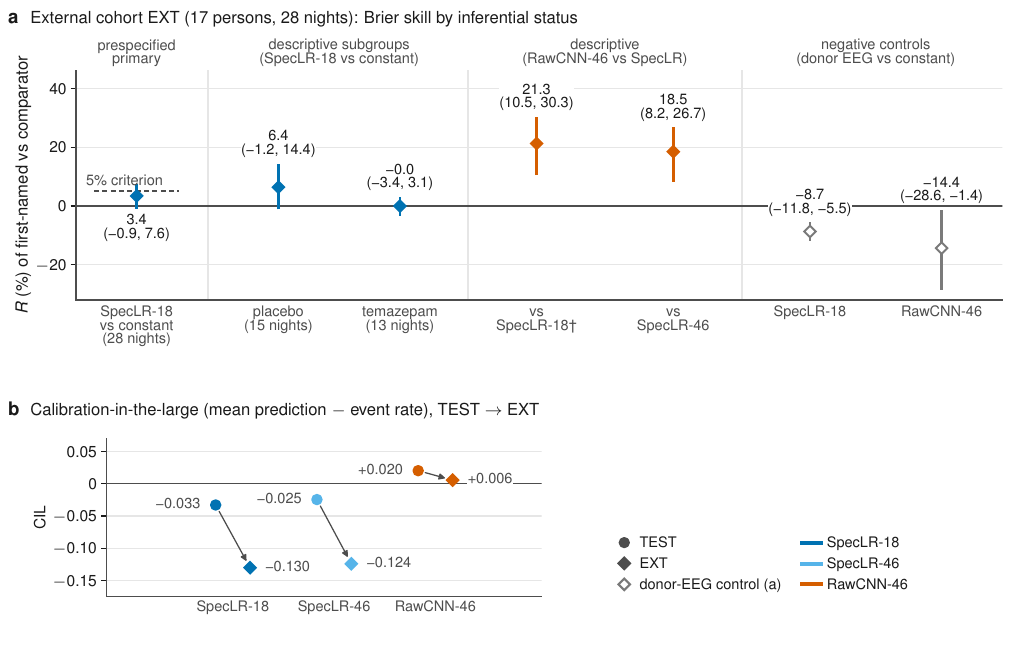}
\caption{External cohort (EXT; 17 persons, 28 nights). (a) Brier skill $R$ with 95\% participant-cluster intervals from the original external evaluation, grouped by inferential status along the horizontal axis: the prespecified primary contrast (short dashed line: its 5\% criterion), descriptive drug-condition subgroups, descriptive EEG-versus-EEG contrasts ($\dagger$: the conditional CNN test was not reached because the primary result was inconclusive) and donor-EEG negative controls (open grey markers; each window's EEG replaced by another person's EEG from the same elapsed-time stratum, not matched on drug condition). Printed values are $R$ (95\% CI). (b) Calibration-in-the-large (mean prediction minus event rate) of each model on TEST (circles) and EXT (diamonds); arrows show the TEST-to-EXT shift. Point values.}
\label{fig:ext}
\end{figure}

\section{Definitions}
\label{app:defs}
\textbf{Risk sets and counts.} Decisions are indexed by person $i$ and night $r$; times are seconds after lights-off, and in every eligible night lights-off and $T_{2,ir}$ fall on the scorer's 30\,s epoch grid. A night's decisions are the indexed tuples $\mathcal{A}_{0,ir}=\{(i,r,t):t\in\{30,60,\dots\},\ t<T_{2,ir}\}$ (the first decision has 30\,s of EEG after lights-off), and $\mathcal{A}_0=\bigcup_{(i,r)\in\mathcal{N}}\mathcal{A}_{0,ir}$, where $\mathcal{N}$ holds one predeclared eligible night per person in the VAL and TEST primary analyses and all eligible nights in EXT, in TRAIN and in the sensitivity analyses. Because each decision carries its person and night, this union is disjoint: decisions at the same elapsed time on different nights remain distinct, and $|\mathcal{A}_0|=\sum_{(i,r)\in\mathcal{N}}|\mathcal{A}_{0,ir}|$. Elsewhere $t$, $Y_t$, $p_t$ and $S(t)$ abbreviate the indexed quantities. Eligibility (App.~\ref{app:cohort}) and $T_{2,ir}$ come from retrospective scoring. The unit of analysis is the decision $(i,r,t)$, and the person is the bootstrap cluster. With $Y_{ir,t}=\mathbf{1}\{t<T_{2,ir}\le t+h\}$ the decision at $t=T_{2,ir}-300$\,s is positive, so a night has $T_{2,ir}/30-1$ decisions and $\min(10,\,T_{2,ir}/30-1)$ positives. Every TEST and EXT night had ten; one VAL primary night had seven ($14\times10+7=147$).

\textbf{Forecast target.} For an EEG forecaster, $p_{ir,t}$ estimates $\pi_{ir,t}=P(T_{2,ir}\le t+h\mid T_{2,ir}>t,\ \mathcal{F}^{\mathrm{EEG}}_{ir,t})$, where $\mathcal{F}^{\mathrm{EEG}}_{ir,t}$ is generated by the two EEG channels on $[t-30,t)$; $Y_{ir,t}$ is the realised outcome. SpecLR and RawCNN receive only that window: no later EEG, scored stage, $T_1$ or $T_2$ is a model input. Membership of the risk set ($T_{2,ir}>t$) and eligibility are known only after scoring and select which decisions are evaluated. The constant uses no covariate; Stage-only and Stage+time condition on the scored stage $S_{ir}(t)$ (and on $t$), which only retrospective scoring provides.

\textbf{Estimands.} E0: $\mathcal{A}_0$ and $Y_t$ as above. E1 (wake$\to$first N1; TRAIN only): $\mathcal{A}_1=\{t<T_1\}$, $Y_t=\mathbf{1}\{T_1\le t+h\}$. E2 (wake input$\to$first N2): $\{t<T_2:S(t)=\mathrm{W}\}$, label as E0, split at first sleep as described next. E3 (N1$\to$N2): $\{t<T_2:S(t)=1\}$, label as E0. E1--E3 use scorer labels to define the risk set and are scorer-defined.

\textbf{Two conventions at first sleep.} The decision at $t=T_1$ uses $[T_1-30,T_1)$, the last all-wake epoch before the first scored sleep epoch. The held-out stage analyses (Tables~\ref{tab:r2full} and~\ref{tab:r3bfull}) call a decision pre-sleep if $t\le T_1$ (its input epoch lies entirely before first sleep) and returning wake if $S(t)=\mathrm{W}$ and $t>T_1$. The TRAIN-only exploratory analysis (App.~\ref{app:w2s}, Table~\ref{tab:w2s}) uses the strict split: E1 and E2-pre require $t<T_1$, and E2-ret is $\{t<T_2:S(t)=\mathrm{W},\ t\ge T_1\}$. Its E2-ret group therefore also contains each night's boundary decision at $t=T_1$ (84 of its 515 windows), whose input is wake before first N1 rather than wake after sleep. Each analysis is reported with the convention it implemented, and no window was reclassified; the TRAIN E2 columns and the held-out pre-sleep and returning-wake rows are not numerically interchangeable. Labels do not depend on either convention.

\textbf{Brier estimators.} With $\ell^{m}_{ir,t}=(p^{m}_{ir,t}-Y_{ir,t})^2$ and $\mathcal{A}_i=\bigcup_{r:(i,r)\in\mathcal{N}}\mathcal{A}_{0,ir}$ the analysed decisions of person $i$, so that $|\mathcal{A}_i|=\sum_{r:(i,r)\in\mathcal{N}}|\mathcal{A}_{0,ir}|$ ($P$ persons), $\dse=\frac{1}{P}\sum_{i=1}^{P}\frac{1}{|\mathcal{A}_i|}\sum_{(i,r,t)\in\mathcal{A}_i}\big(\ell^{\mathrm{ref}}_{ir,t}-\ell^{\mathrm{c}}_{ir,t}\big)$: a person's nights (one in the VAL and TEST primary analyses, one or two in EXT) are pooled with window weighting, and persons are then weighted equally; the subject-equal Brier of one model averages $|\mathcal{A}_i|^{-1}\sum_{(i,r,t)\in\mathcal{A}_i}\ell^{m}_{ir,t}$ over persons in the same way. $\R=\dww/\bar B_{\mathrm{ref}}$ is positive when the candidate improves on its reference. Because $\bar B_{\mathrm{ref}}$ is reference-specific, skills relative to different references cannot be subtracted, whereas on identical decisions $\dww$ is additive: for forecasts $a$, $b$ and $c$, $\dww(a\to c)=\dww(a\to b)+\dww(b\to c)$.

\textbf{Stage-only reference.} $\hat p_s=(k_s+1)/(n_s+2)$, where $k_s$ of the $n_s$ TRAIN decisions with scored input stage $s$ are positive (46 persons, all 84 nights): $(249+1)/(3{,}573+2)=0.070$ for W and $(582+1)/(1{,}219+2)=0.477$ for N1. The 18-person sensitivity refit uses the same estimator. Like Stage+time, it requires the retrospectively scored stage and is a diagnostic, not a deployable forecaster.

\textbf{SpecLR features and fit.} Each decision window holds 3{,}000 samples per channel (30\,s at 100\,Hz). For each channel, five non-overlapping 4\,s Hann-tapered frames starting 0, 6, 12, 18 and 24\,s into the window (samples 0--399, 600--999, 1{,}200--1{,}599, 1{,}800--2{,}199 and 2{,}400--2{,}799; 400 samples zero-padded to a 512-point FFT) give, for 0.5--4, 4--8, 8--13 and 13--30\,Hz, the sum of squared FFT magnitudes over the five frames and the in-band bins; features are $\log(1+\text{summed power})$ in each band plus $\log(1+\text{AC RMS})$ computed over all 3{,}000 samples, i.e.\ ten features over two channels. The band powers therefore sample 20\,s of each 30\,s window, whereas the RMS uses the full window; this frame layout was fixed in the TRAIN-only feature pilot before any held-out use, and its original motivation was not documented. SpecLR-18 and SpecLR-46 share this recipe and estimator, RawCNN-46 receives all $2\times3{,}000$ raw samples, and none of the three receives elapsed time. Features are z-scored with training means and population standard deviations (floor $10^{-6}$) and clipped to $\pm5$. Logistic regression is fitted by 350 steps of gradient descent (learning rate $0.3/(1+\text{step}/200)$, L2 penalty 0.025); predictions are clipped to $[10^{-6},1-10^{-6}]$.

\textbf{RawCNN architecture and training.} After per-channel mean subtraction, each 30\,s input (2 channels, 3{,}000 samples) is divided by the fixed training scale of 10.141\,\textmu V and clipped to $\pm10$. Three parallel 1D convolutional stems each map 2 channels to 32 (kernels 7, 25, 101; stride 2; symmetric padding), and their 96-channel concatenation passes through GELU and GroupNorm (4 groups). The body has three kernel-7, stride-4, padding-3 convolutions: 96$\to$64, 64$\to$64 and 64$\to$64 channels. GELU follows each, with 4-group GroupNorm after the first two. Global temporal mean and $\log(\mathrm{variance}+10^{-6})$ are concatenated (128 dimensions), projected through Linear(128,32)+GELU, then Linear(32,32)+GELU, and finally Linear(32,1) to a binary logit (114{,}817 trainable parameters). The raw-only model has no trajectory GRU, elapsed-time features or hazard head. The output-head bias is initialized at the training prevalence, and optimization uses binary cross-entropy with logits, AdamW (learning rate $10^{-3}$, weight decay $10^{-3}$), batch size 256 and gradient-norm clipping at 1. Final fits use 4 epochs (chosen from development CV); probabilities are averaged over three seeds. The archived evaluation code documents the model's layer order, parameter count, and final-fit settings.

\textbf{Calibration and discrimination.} Calibration-in-the-large is mean prediction minus event rate. Calibration slope is the coefficient of a logistic regression of $Y$ on $\mathrm{logit}(p)$ (essentially unpenalised). Within-recording AUROC averages AUROC computed within each recording. The retrospective \emph{exact-landmark-matched} AUROC compares positives with negatives from other persons at the same 30\,s landmark index $\lfloor t/30\rfloor$. The original held-out \emph{elapsed-stratum-matched} AUROC instead compares within coarser elapsed-time bins (Table~4). These estimators are not interchangeable. References constant within each compared stratum obtain AUROC 0.5 by construction.

\textbf{Stage-reference selection.} Configurations: spline knots $\{3,5\}$, $C\in\{0.1,1,10\}$, with or without a stage-by-time interaction; selection by summed Brier over person-disjoint inner folds of TRAIN. The 18-person refit selected 3 knots and $C=1$.

\clearpage
\section{Contribution versus prior art}
\label{app:priorart}
{\footnotesize
\setlength{\tabcolsep}{3pt}
\begin{longtable}{@{}p{0.27\linewidth}p{0.38\linewidth}p{0.29\linewidth}@{}}
\caption{Contribution versus prior art: each component, its established prior art and its status in this paper.}\label{tab:priorart}\\
\toprule
Component & Established prior art & Status here \\
\midrule
\endfirsthead
\toprule
Component & Established prior art & Status here \\
\midrule
\endhead
Brier score, skill, decomposition; proper scoring rules & \citet{brier1950,murphy1973partition,gneiting2007scoring} & Used; not novel \\
Conditional at-risk predictions at successive times & Landmarking \citep{vanhouwelingen2007landmark}; time-dependent Brier \citep{graf1999assessment} & E0--E3 are landmark-style targets \\
State-dependent transition probabilities & Multistate models \citep{putter2007multistate} & Stage references are minimal state-conditional landmark models \\
Calibration as outcome; shift degrades calibration & \citet{vancalster2019calibration,ovadia2019trust,collins2024tripodai} & Applied \\
Evaluation of EEG event prediction & \citet{winterhalder2003spc,proix2021seizure} & Warning analysis follows the same logic \\
Pooled cross-fold AUROC bias & \citet{forman2010apples} & Encountered in the exploratory analysis; not novel \\
EEG approach to first N2; latency prediction & \citet{li2025falling,su2026presleep,beppu2025latency,oh2024medisol,kweon2026ssfset} & Different alignment, target or metrics \\
\textbf{Stage attribution and state-conditioned comparisons on previously held-out predictions} & Established statistical ingredients; combination evaluated in this case study (not a priority claim) & Evaluative contribution \\
\textbf{Empirical findings} & --- & Specific to Sleep-EDF; generality untested \\
\bottomrule
\end{longtable}}

\FloatBarrier
\section{Stage-stratified evaluation of held-out predictions}
\label{app:state_strata}
\TableRtwoFull

\section{Held-out comparisons with scored-stage references}
\label{app:stage_reference}
\TableRthreeB

\section{Exploratory wake-to-first-N1 programme}
\label{app:w2s}
\textbf{Notes.} Estimands, references and model families were fixed in writing before fitting. On the pooled 300\,s matching, the time-only reference scored AUROC 0.377, because each fold's reference is fitted on the other folds and its level is anti-ordered with the fold's own event rate (fold-level correlation $-0.95$ for the constant); this is an instance of the cross-fold pooling bias described by \citet{forman2010apples}. With exact 30\,s matching within folds the time reference scored 0.500 by construction and the best EEG model 0.730\ci{0.646}{0.801}. E2-pre and E2-ret follow the strict TRAIN split at first sleep ($t<T_1$ versus $t\ge T_1$; App.~\ref{app:defs}), so E2-ret also contains each night's last all-wake decision before first N1, and neither column equals a held-out pre-sleep or returning-wake stratum. On E2 pre-sleep inputs the 30\,s CNN was worse than time ($-$40.8\%\ci{$-$101.7}{$-$3.9}); on E3 a spectrogram model improved on time by 20.6\%\ci{8.1}{31.2}. Recalibration raised the best model from 0.1\%\ci{$-$20.6}{17.7} to 7.8\%\ci{$-$0.2}{14.6} and moved its calibration-in-the-large from $+0.051$ to 0.000. Replacing a person's history by another person's left predictions essentially unchanged (own minus donor 0.0\%\ci{$-$0.3}{0.4}).

\begin{figure}[H]
\centering
\includegraphics[width=\textwidth]{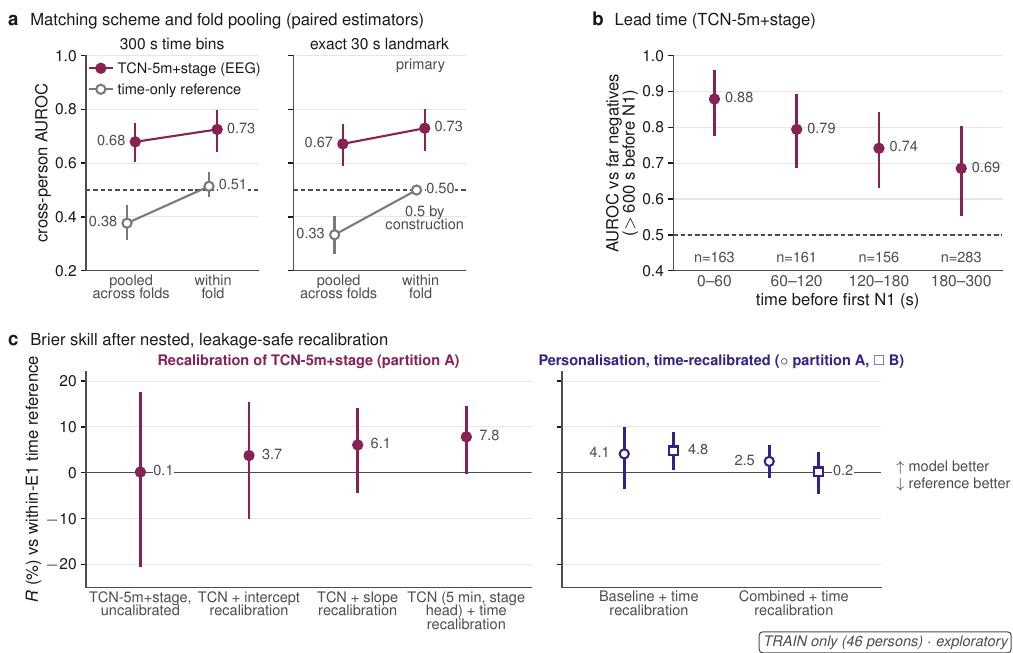}
\caption{Exploratory TRAIN-only analysis of E1 (wake$\to$first N1; 3{,}058 windows, 763 positive, 46 persons), with 95\% person-cluster bootstrap intervals. (a) Cross-person AUROC of the TCN-5m+stage model (filled) and the time-only reference (open grey) under 300\,s time-bin or exact 30\,s landmark matching (one facet each). A line joins the same model's estimates with pairs pooled across folds and with pairs restricted to the same fold: two estimators on the same windows, not two time points. Under the primary within-fold exact matching the time reference is 0.5 by construction. (b) AUROC of positive windows in each lead-time bin against far negatives (more than 600\,s before first N1), using within-fold exact-landmark pairs; $n$ = positive windows. The bins are not connected because their overlapping intervals do not establish a monotonic trend. (c) Brier skill relative to a within-E1 time reference after nested, leakage-safe recalibration: recalibration steps of TCN-5m+stage (left; unconnected categories) and personalised variants on two fold partitions (right; open markers, circles: partition A, squares: partition B); labels as in Table~\ref{tab:p0p1}. One colour per model family is used in all panels.}
\label{fig:w2s}
\end{figure}
\TableWtwoS
\TablePzeroPone

\section{Architecture ablation study (development cross-validation)}
\label{app:ablation}
\begin{figure}[H]
\centering
\includegraphics[width=\textwidth]{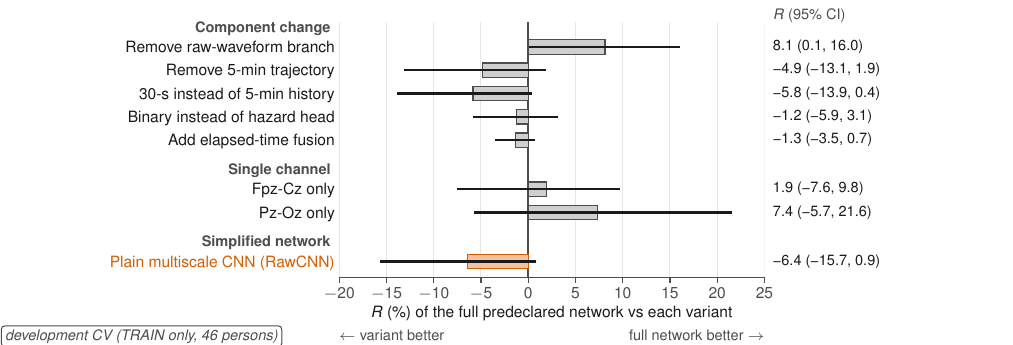}
\caption{Predeclared task-specific network versus each variant (5-fold TRAIN CV, 46 persons). Bars show $R=(B_{\mathrm{variant}}-B_{\mathrm{full}})/B_{\mathrm{variant}}$ from zero and whiskers its 95\% person-cluster CI; $R>0$ favours the full predeclared network over the displayed variant; for a removal, it means retaining that component improved performance. The highlighted plain CNN became RawCNN, a generic, not a proposed, architecture.}
\label{fig:ablation}
\end{figure}
{\widowpenalty=10000 In Fig.~\ref{fig:ablation}, removing the raw-waveform branch worsened Brier ($+8.1\%$\ci{0.1}{16.0}); removing the 5-min trajectory, replacing the hazard head, or adding elapsed-time fusion did not help, and the plain CNN was not worse than the full network ($-6.4\%$\ci{$-$15.7}{0.9}).\par}

\section{Representation and perturbation analysis (development data)}
\label{app:interp}
\begin{figure}[H]
\centering
\includegraphics[width=\textwidth]{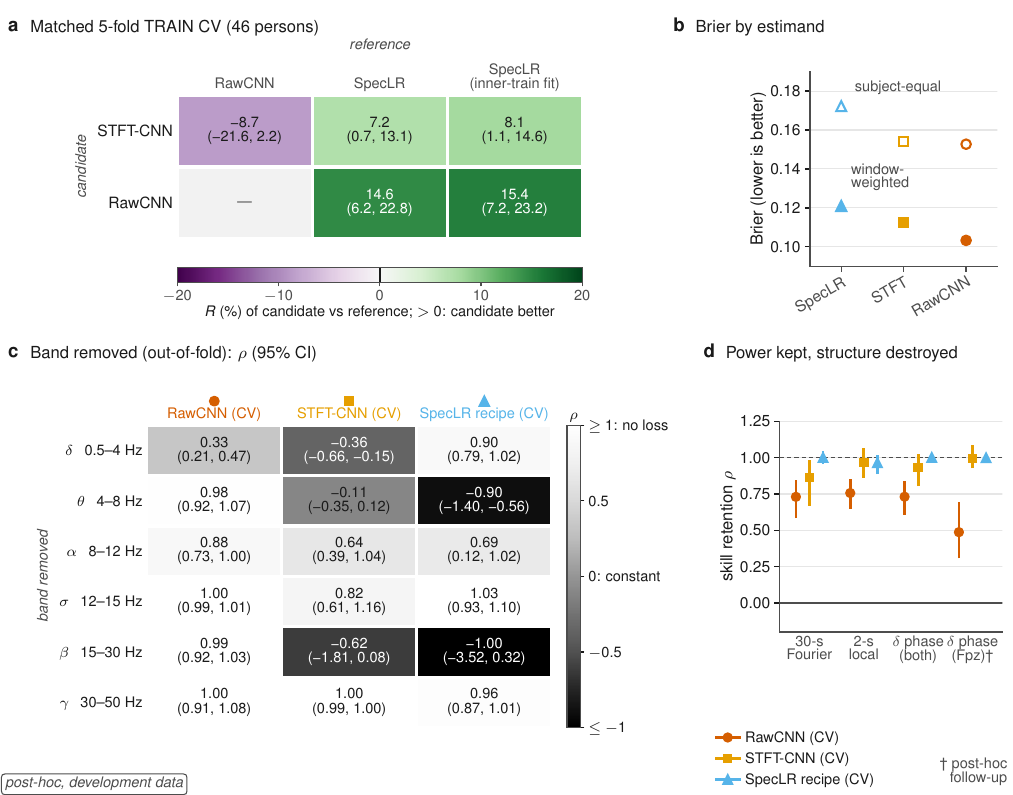}
\caption{Post-hoc analyses on development data (5-fold TRAIN CV, 46 persons; 95\% person-cluster intervals). (a) $R$ (\%) of the candidate (row) relative to the reference (column), printed as $R$ (95\% CI) and shaded on a diverging scale centred at zero (green: candidate better); the diagonal cell is empty. (b) Window-weighted (filled) and subject-equal (open) Brier. (c) Skill retention $\rho$ (fraction of skill over the constant retained; 1 = unchanged, $>1$ = improved relative skill, $<0$ = worse than the constant) after removing one band, printed as $\rho$ (95\% CI) in each cell and shaded from $\rho\le-1$ (black) to $\rho\ge1$ (white); shading saturates only above 1 (at most 1.03), printed values are exact. (d) Retention after power-preserving, structure-destroying perturbations ($\dagger$ post-hoc follow-up). Marker shape repeats colour. Sensitivity results only; no mechanism or representation is established.}
\label{fig:interp}
\end{figure}

A capacity-matched log-power STFT CNN was not distinguishable from the raw CNN ($\R=-8.7\%$\ci{$-$21.6}{2.2}). Removing 0.5--4\,Hz content reduced the raw CNN's skill retention to 0.33\ci{0.21}{0.47} against 0.97 for an energy-matched control; the spectral model was most sensitive to 4--8 and 15--30\,Hz power. Integrated Gradients \citep{sundararajan2017ig} failed model-randomisation sanity checks \citep{adebayo2018sanity}, and a deletion test \citep[cf.][]{hooker2019roar} did not beat ranking by input magnitude; we draw no conclusions from attributions. Surrogates follow \citet{theiler1992surrogate,prichard1994surrogate}. These are sensitivity results (Fig.~\ref{fig:interp}) and establish no physiological mechanism.

\section{Statistical units, bootstrap procedure and retrospective supplementary contrasts}
\label{app:stat_supplement}
\textbf{Independent units.} Decisions within a night share one event time and are serially dependent, and EXT includes two nights for 11 of its 17 persons. The independent sampling units are therefore persons: 15 (VAL), 15 (TEST) and 17 (EXT), of whom 13, 14 and 17 contribute N1-scored decisions. Hundreds of decisions per cohort do not add independent persons.

\textbf{Bootstrap procedure.} Each replicate draws persons with replacement, as many as there are persons in the cohort or stratum; a drawn person contributes every analysed decision of every analysed night. Pooled quantities ($\dww$, $\R$, $C_s$) are recomputed from the resampled sums of losses and decision counts, and person-equal quantities ($\dse$) average the drawn persons' mean differences. Stratum intervals resample only the persons with decisions in that stratum and are not reported for fewer than eight persons. Models and references are frozen and never refitted. Intervals are 95\% percentiles of 10{,}000 replicates; prespecified results keep their original seeds and retrospective analyses use new ones. Replicates add no independent persons, and percentile intervals with 13--17 clusters are approximate.

\textbf{Retrospective between-cohort contrasts.} A within-stratum advantage observed in one cohort but not in another does not by itself show a difference between cohorts. We therefore estimated three direct contrasts on the same frozen candidate, the same TRAIN-fitted Stage+time reference and the same scored input stratum, resampling persons independently within each cohort (Table~\ref{tab:cohort_contrasts}). These contrasts were selected after the per-cohort results had been seen; their procedure, seeds and intervals were fixed and recorded before the computation, but the questions are retrospective and the results descriptive. Bonferroni-adjusted intervals for the family of three serve as a multiplicity sensitivity analysis: they exclude zero for D1 and, narrowly, for D2b (lower limit 0.00038), and include zero for D2a. The cohorts also differ in study setting, drug condition, N1 prevalence, decisions per person and selection, so a non-zero $D$ describes a difference in recorded improvement and does not identify its cause; an interval including zero would not establish equality. TRAIN cross-validation uses fold-refitted models and references and is not used as a comparison cohort.

\begin{table}[H]
\centering
\caption{Retrospective direct between-cohort contrasts of absolute Brier improvement over Stage+time within a scored input stratum, $D=\dww^{(A)}-\dww^{(B)}$ (same frozen candidate and TRAIN-fitted reference in both cohorts; persons resampled independently within each cohort, 10{,}000 replicates). Persons: independent persons with decisions in the stratum (A / B). Bonferroni: 98.33\% interval for the family of three contrasts. $D_{\mathrm{SE}}$: the same contrast of person-equal improvements (95\% interval). Differences of $R$ are not shown because the reference Brier scores differ between cohorts. Descriptive; selected after the per-cohort results had been seen.}
\label{tab:cohort_contrasts}
\footnotesize
\setlength{\tabcolsep}{3pt}
\resizebox{\linewidth}{!}{%
\begin{tabular}{@{}llcccclll@{}}
\toprule
Contrast & Candidate, stratum & Persons & $\dww^{(A)}$ & $\dww^{(B)}$ & $D$ & 95\% & Bonferroni & $D_{\mathrm{SE}}$ (95\%) \\
\midrule
D1: TEST $-$ EXT & RawCNN-46, N1 & 14 / 17 & 0.0648 & $-$0.0365 & 0.1013 & (0.0372, 0.1645) & (0.0231, 0.1784) & 0.1314 (0.0551, 0.2141) \\
D2a: VAL $-$ TEST & SpecLR-18, W & 15 / 15 & 0.0108 & $-$0.0042 & 0.0151 & (0.0017, 0.0411) & ($-$0.0004, 0.0475) & 0.0417 (0.0051, 0.0899) \\
D2b: VAL $-$ EXT & SpecLR-18, W & 15 / 17 & 0.0108 & 0.0004 & 0.0104 & (0.0018, 0.0362) & (0.0004, 0.0425) & 0.0460 (0.0090, 0.0936) \\
\bottomrule
\end{tabular}}
\end{table}

\textbf{Uncertainty of stage attribution.} For the seven attributions of Fig.~\ref{fig:attr}, Table~\ref{tab:attr_ci} gives person-cluster intervals for $\dww$, $C_{\mathrm{W}}$ and $C_1$; the pooled denominator is resampled with the numerators, and $\dww=C_{\mathrm{W}}+C_1$ holds in every replicate. The share $Q_1$ receives an interval only when its denominator is stable: at least eight persons, fewer than 0.5\% of replicates with $\dww$ of opposite sign or zero, and fewer than 2.5\% with $|\dww|$ below a quarter of its point value. This is a descriptive heuristic, not a coverage guarantee. The EXT SpecLR-18 denominator fails it (6.0\% and 9.0\%), so its point share of 35.1\% is unreliable and is not interpreted. Shares can exceed 100\% or be negative when the other stage contributes negatively; they are accounting quantities, not causal or physiological fractions, and an interval for $C_{\mathrm{W}}$ that includes zero does not show that the wake contribution is zero.

\begin{table}[H]
\centering
\caption{Uncertainty of the stage attribution of Fig.~\ref{fig:attr}: pooled improvement over the constant, wake and N1 contributions (absolute Brier units) and N1 share $Q_1=C_1/\dww$ (\%), with 95\% person-cluster intervals (10{,}000 replicates; every drawn person contributes all its W and N1 decisions). Share intervals are shown only when the denominator is stable (see text). $^{*}$The denominator changed sign or fell below a quarter of its point value in 6.0\% and 9.0\% of replicates; the point share is not interpretable.}
\label{tab:attr_ci}
\footnotesize
\setlength{\tabcolsep}{4pt}
\resizebox{\linewidth}{!}{%
\begin{tabular}{@{}llclllc@{}}
\toprule
Set & Model & Persons & $\dww$ (95\%) & $C_{\mathrm{W}}$ (95\%) & $C_1$ (95\%) & $Q_1$ (95\%) \\
\midrule
VAL & SpecLR-18 & 15 & 0.0305 (0.0189, 0.0560) & 0.0114 (0.0040, 0.0189) & 0.0191 (0.0077, 0.0434) & 62.7 (38.1, 89.2) \\
TEST & SpecLR-18 & 15 & 0.0280 (0.0115, 0.0568) & 0.0004 ($-$0.0048, 0.0055) & 0.0277 (0.0126, 0.0554) & 98.6 (81.2, 128.7) \\
TEST & SpecLR-46 & 15 & 0.0353 (0.0174, 0.0678) & 0.0005 ($-$0.0045, 0.0054) & 0.0348 (0.0186, 0.0661) & 98.5 (86.7, 118.0) \\
TEST & RawCNN-46 & 15 & 0.0578 (0.0201, 0.1195) & $-$0.0010 ($-$0.0158, 0.0149) & 0.0588 (0.0314, 0.1098) & 101.8 (82.3, 169.8) \\
EXT & SpecLR-18 & 17 & 0.0055 ($-$0.0017, 0.0110) & 0.0035 ($-$0.0022, 0.0072) & 0.0019 ($-$0.0022, 0.0067) & 35.1, unreliable$^{*}$ \\
EXT & SpecLR-46 & 17 & 0.0107 (0.0044, 0.0161) & 0.0039 ($-$0.0016, 0.0074) & 0.0068 (0.0028, 0.0122) & 63.4 (36.1, 124.0) \\
EXT & RawCNN-46 & 17 & 0.0383 (0.0208, 0.0621) & $-$0.0016 ($-$0.0062, 0.0028) & 0.0399 (0.0233, 0.0639) & 104.1 (93.3, 120.3) \\
\bottomrule
\end{tabular}}
\end{table}

\section{Additional uncertainty details for secondary contrasts}
\label{app:secondary_intervals}
This section provides interval estimates for secondary and exploratory contrasts not tabulated elsewhere. The reported intervals retain their original analysis-specific definitions and are descriptive unless explicitly identified as belonging to a prespecified test. Interval overlap alone is not a formal test of differences.

\textbf{Prespecified discrimination.} SpecLR-18's within-recording AUROC was 0.82\ci{0.76}{0.86} in EXT and 0.80\ci{0.67}{0.91} in TEST; this was a descriptive, not formally tested, comparison.

\textbf{TRAIN exploratory and reference comparisons.} The CNN recipe's wake-only skill was 20.3\%\ci{8.9}{30.5}. Two time-only references had $-0.7$\%\ci{$-$2.9}{1.4} and 0.5\%\ci{$-$1.8}{2.9} relative skill in cross-validation. Stage-only improved on constant by 22.0\%\ci{14.8}{28.8}. Relative to Stage-only, the spectral recipe had $-7.4$\%\ci{$-$20.6}{3.5} and the CNN recipe 8.3\%\ci{$-$2.3}{18.1}. The CNN's N1-specific skill beyond Stage+time in TRAIN cross-validation was 10.6\%\ci{$-$3.3}{21.8}.

\textbf{Exact-landmark discrimination.} Within wake-scored inputs, RawCNN-46's cross-person AUROC was 0.80\ci{0.65}{0.95} in TEST and 0.60\ci{0.49}{0.72} in EXT; SpecLR-18 was 0.78\ci{0.62}{0.91} in VAL. Within N1 inputs, RawCNN-46 reached 0.64\ci{0.43}{0.88} in EXT. Over all input stages, Stage-only yielded 0.79\ci{0.70}{0.89} in TEST and 0.76\ci{0.69}{0.83} in EXT.

\textbf{Calibration decomposition.} The between-person probability-level component of SpecLR-18's Brier difference in EXT was $-0.012$\ci{$-0.021$}{$-0.006$}; this is a prespecified secondary, descriptive comparison.

\textbf{Wake-to-first-N1 discrimination (TRAIN only; exploratory).} Time-matched cross-person AUROC as reported in the frozen exploratory programme, stratified by time before first N1, was 0.88\ci{0.77}{0.96} within 60\,s, 0.79\ci{0.69}{0.89} at 1--2\,min, 0.74\ci{0.63}{0.84} at 2--3\,min and 0.69\ci{0.55}{0.80} at 3--5\,min. The intervals overlap and do not establish monotonic change.

\textbf{Warning utility (TRAIN only; exploratory).} For the recalibrated first-N1 model, at a fixed, untuned threshold, recalibration reduced false-alarm episodes from 3.85 to 0.71 per hour but also reduced the proportion of first-N1 onsets preceded by an alarm within the five-minute horizon from 59\% to 24\%; recent-EEG personalization offered no additional benefit (Table~\ref{tab:p0p1}).

\section{Reproducibility, compute and LLM usage}
\label{app:repro}
\textbf{Reproducibility status.} The implementation and validation procedures are available in a separate MIT-licensed code repository. The code was evaluated using synthetic tests and TRAIN-only method-equivalence checks. \textbf{Compute.} All analyses were conducted on a single workstation equipped with a 10-core CPU and an NVIDIA RTX 5090 GPU with 32\,GB of memory. Confirmatory statistical tests were performed on the CPU. \textbf{Use of language models.} AI-based language tools assisted with manuscript preparation. The authors take full responsibility for the manuscript content, code, and data handling.

\FloatBarrier
\widowpenalty=10000
\section{Code and data availability}
\label{app:code_access}
The source code for the methods and analyses reported in this paper is maintained in a code repository under the MIT software licence. The package includes executable implementations of cohort preparation, SpecLR and RawCNN, held-out evaluation logic, Stage-only and Stage+time diagnostic references, and TRAIN-only exploratory, ablation and perturbation analyses. Sleep-EDF Expanded is obtained independently from PhysioNet under its own licence; no raw EEG, split rosters, person-level prediction files, held-out model weights, or stage-join artifacts are redistributed.
The public source-code repository is available at \url{https://github.com/AliAmini93/h-eval-sleep-onset-eeg}.

\end{document}